\documentclass[useAMS,usenatbib,usegraphicx]{mn2e}

\usepackage{color, aas_macros, pifont}
\def\eref#1{equation (\ref{eq:#1})}
\def\eeref#1{(\ref{eq:#1})}
\def\fref#1{Fig. \ref{fig:#1}}
\def\sref#1{section \ref{sec:#1}}

\def\unitx{\hat{\bmath{e}}_x}
\def\unity{\hat{\bmath{e}}_y}
\def\unitz{\hat{\bmath{e}}_z}

\def\bcdot{\bmath{\cdot}}

\def\change#1{{#1}}
\def\cchange#1{{#1}}

\def\p1{p^{(1)}}
\def\drho1{\rho^{(1)}}
\def\B1{\bmath{B}^{(1)}}
\def\V1{\bmath{v}^{(1)}}
\def\b1{B^{(1)}}
\def\dv1{v^{(1)}}
\def\a1{A^{(1)}}
\def\i{\mathrm{i}}

\def\flone{\ding{192}}
\def\fltwo{\ding{193}}
\def\flthree{\ding{194}}
\def\flfour{\ding{195}}

\title{Resistive relaxation of a magnetically confined mountain on an
  accreting neutron star}

\author[M. Vigelius et al.]{M.~Vigelius $^1$\thanks{E-mail: mvigeliu@physics.unimelb.edu.au} and A.~Melatos $^1$ \\
 $^1$ School of Physics, University of Melbourne, Parkville, VIC
 3010, Australia}

\begin{document}

\date{Submitted to MNRAS}

\maketitle

\begin{abstract}
  Three-dimensional numerical magnetohydrodynamic (MHD) simulations
  are performed to investigate how a magnetically confined mountain
  on an accreting neutron star relaxes resistively.
  No evidence is found for non-ideal MHD instabilities on a short
  time-scale, such as the resistive ballooning mode or the tearing
  mode. Instead, the mountain relaxes gradually as matter is
  transported across magnetic surfaces on the diffusion time-scale,
  which evaluates to \change{$\tau_\mathrm{I} \sim 10^5 - 10^8$} yr
  (\cchange{depending on the conductivity of the neutron star crust}) for an accreted
  mass of $M_a = 1.2 \times 10^{-4} M_\odot$. The magnetic dipole moment
  simultaneously reemerges as the screening currents dissipate over
  $\tau_\mathrm{I}$. For nonaxisymmetric mountains, ohmic dissipation
  tends to restore axisymmetry by magnetic reconnection at a
  filamentary neutral sheet in the equatorial  plane. Ideal-MHD oscillations on the 
  Alfv\'{e}n time-scale, which can be excited by external influences, such
  as variations in the accretion torque, compress the magnetic field
  and hence decrease $\tau_\mathrm{I}$ by \change{one order of magnitude}
  relative to its standard value (as computed for the static
  configuration). The implications of long-lived mountains for
  gravitational wave emission from low-mass X-ray binaries are briefly
  explored.
\end{abstract}

\bibliographystyle{mn2e}


\section{Introduction}
Observations suggest that the
magnetic dipole moment of accreting neutron stars in X-ray
binaries, $\mu$, decreases with accreted mass, $M_a$
\citep{Taam86, vanDenHeuvel95}, possibly through magnetic screening or
burial \citep{Bisnovatyi74,Romani90,Payne04,Lovelace2005}.
During the burial process, the accreted plasma is channeled onto the
magnetic poles of the neutron star, whence it spreads
equatorwards, thereby distorting the frozen-in magnetic flux
\citep{Melatos01}. Quasistatic sequences of ideal-magnetohydrodynamic (ideal-MHD)
equilibria describing how burial proceeds were computed by
\citet{Payne04} (hereafter PM04). These authors
found that the magnetic field is compressed into an equatorial belt,
which confines the accreted mountain at the poles.

Surprisingly, the distorted equilibrium magnetic fields resulting from
burial are stable in ideal MHD. In an axisymmetric analysis,
\citet{Payne07} (hereafter PM07) found that the mountain, when perturbed,
oscillates radially and laterally in a superposition of global Alfv\'{e}n and
compressional modes, but it remains intact. Of course, an axisymmetric
analysis neglects important toroidal modes. \citet{Vigelius08a} (hereafter
VM08) found that the axisymmetric configuration is unstable to
the undulating submode of the three-dimensional Parker instability in
spherical geometry. Again though, while the hydromagnetic structure
reconfigures itself globally, the mountain remains confined to the magnetic
poles once the instability saturates.

PM04, PM07, and VM08 considered ideal-MHD equilibria. However,
magnetic burial creates steep magnetic gradients, which relax
resistively. A conservative estimate of the
relative importance of nonideal effects can be 
arrived at by assuming that the electrical resistivity in the outer
crust is dominated by electron-phonon scattering. Under this
assumption, resistive relaxation arrests the growth of a mountain
when the accreted mass exceeds $\sim 10^{-5}
M_\odot$ \citep{Brown98, Cumming04,Melatos05}.
Resistive instabilities, like the
global tearing mode or the local gravitational mode \citep{Furth63},
grow faster than the simple resistive
time-scale. Three-dimensional modes
like the resistive ballooning mode, which grows if the pressure
gradient is parallel to the field line curvature, may rapidly destroy the
confinement of the mountain.

\change{
During the early stages of accretion, the mountain might
be disrupted on the Alfv\'{e}n time scale by the ideal-MHD ballooning
mode \citep{Litwin01}. \citet{Vigelius08b} and \citet{Vigelius08e}
show that equatorial magnetic stresses stabilize the configuration and generally prevent disruption
in the high-$M_a$ regime. These authors then continue to solve the
initial value problem by injecting plasma into an initially
homogenuous background (with $M_a=0$) threaded by a dipolar
field. They find no evidence for a growing instability in the
low-$M_a$ regime. In this article, we investigate further how resistive
relaxation competes with accretion at different accretion rates.
}

The main aim of this article is to test if resistive instabilities
disrupt the mountain on time-scales comparable to the accretion 
time-scale.
The article is divided into six sections. Section \ref{sec:model} introduces the
numerical setup used in our simulations,
\sref{resistive_instabilities} describes the dynamics of the resistive relaxation, and
\sref{field_structure} characterizes the magnetic field structure. In
\sref{relaxation_time}, we evaluate the resistive relaxation time as a
function of accretion parameters. In \sref{reemergence}, we study how
rapidly the magnetic field reemerges after accretion stops. We discuss
our results in \sref{discussion}, focussing on the
ramifications for gravitational wave emission from accreting neutron
stars.

\section{Numerical model}
\label{sec:model}

\subsection{Grid and units}
\label{sec:model:units}
The simulations in this paper employ the parallel, ideal-MHD solver \textsc{zeus-mp}
\citep{Hayes06}, extended to include resistive effects, as described
in appendix \ref{sec:app:resistivity}. All the simulations are carried out in a
spherical polar coordinate system $(r, \theta, \phi)$, where $r$ is
logarithmically stretched as described in PM07 and
VM08. To handle the disparate radial and lateral length-scales, we set
up a downscaled neutron star with $M_\ast=1.01\times10^{-5} M_\odot$
and $R_\ast=2.7\times10^3$ cm, such that the curvature $a=R_\ast/h_0=50$ is still
large while the hydrostatic scale height $h_0=53.8$ cm (defined in
PM04) is preserved. We justify this approach by noting that the small-$M_a$ analytic solution 
depends on $M_\ast$ and $R_\ast$ only through the combination $h_0$ (PM04). The downscaling transformation was
employed in \citet{Payne07} and VM08 and validated by
\citet{Vigelius08b} in the large-$M_a$ regime.

Throughout this paper we fix $\mu_0=G=c_s=h_0=1$, such that the
base units (in cgs) for mass, magnetic field, time, and resistivity become $M_0=h_0
c_s^2/G=8.1\times10^{24}$ g, $B_0=[\change{\mu_0} c_s^4/(G
h_0^2)]^{1/2}=7.2\times10^{17}$ G, $\tau_0=h_0/c_s=5.4\times
10^{-7}$ s, and $\eta_0=\tau_0^{-1}=1.86 \times 10^6$ s$^{-1}$
respectively. The characteristic mass (PM04) then
evaluates to $M_c=6.2 \times 10^{-15} M_\odot$ for the downscaled star.

\subsection{Initial and boundary conditions}
\label{sec:model:conditions}
Our aim in this paper is to examine the influence of a finite
conductivity on magnetic mountain equilibria in 2.5 and 3 dimensions. Axisymmetric
equilibria are imported from the Grad-Shafranov (GS) solver developed by
PM04. Nonaxisymmetric equilibria are imported from
\textsc{zeus-mp} after the transient, three-dimensional Parker instability
saturates (VM08). All our simulations
are isothermal (\texttt{XISO=.true.}).

Boundary conditions are enforced in \textsc{zeus-mp} by ghost cells
framing the active grid. Our choice of a spherical polar grid requires
periodic boundary conditions at the $\phi$ boundaries
[\texttt{ikb.niks(1)=4} and \texttt{ikb.noks(1)=4}]. The
$\theta=\pi/2$ boundary is reflecting [\texttt{ojb.nojs(1)=5}], with
$\bmath{v}_\perp=\bmath{B}_\parallel=0$. The line $\theta=0$
is also reflecting [\texttt{ijb.nijs(1)= -1}], with tangential magnetic field
($\bmath{v}_\perp=\bmath{B}_\perp=0$). Additionally, the toroidal
component $B_\phi$ reverses at $\theta=0$,
i.e. $B_\phi(-\theta)=B_\phi(\theta)$. The
outer boundary at $r=R_m$  is a zero-gradient boundary [\texttt{oib.nois(1)=
  2}]. The magnetic field at $r=R_\ast$ is line-tied by
fixing the plasma variables [\texttt{iib.niis(1)= 3}] at
this boundary: $\bmath{B}$ is dipolar and $\rho$ is kept several
orders of magnitudes higher than the active grid values in order to
realise an impenetrable surface.

\subsection{Resistivity}
\label{sec:model:resistivity}
\change{The electrical conductivity $\sigma$ is a key input into the models presented
in this article. In the outer crust, all transport processes are dominated
by electrons scattering off phonons and impurities [for a recent
review compare \citet{Chamel08}] and $\sigma$ can be derived from
the scattering frequencies in the relaxation time approximation. For
temperatures below the Umklapp temperature ($T_\mathrm{U} \approx 10^7$ K), 
electron-phonon scattering is suppressed and the conductivity must be
attributed to impurities \citep{Cumming01, Cumming04}. In rapid accretors ($\dot{M} \ga 10^{-11}
M_\odot$ yr$^{-1}$), one finds $T \ga 10^8$ K and phonon
scattering dominates, provided the impurity concentration satisfies $Q \la
1$. In accreting neutron stars, $Q$ is set 
by the composition of the ashes produced in steady state
nuclear burning at low densities. \citet{Schatz99} find a large variety
of nuclei in the crust so the impurity factor
is high ($Q \approx 100$). They argue that impurity scattering therefore
dominates, except for very rapid accretors ($Q \sim 1$
for $\dot{M} \ga 30 \dot{M}_\mathrm{Edd}$). On the other hand, \citet{Jones04} noted that,
if the primordial crust is completely replaced by heterogeneous
accreted matter, a temperature-independent
conductivity dominates electron scattering and one finds $Q \gg 1$.
Most authors \citep{Konar97, Cumming01, Cumming04, Pons07} assume $Q
\ll 1$, as do we.}

\change{Neglecting impurities, \citet{Potekhin99a} compute the
  frequency of electron-ion scattering in liquid and solid Fe matter
  for a variety of temperatures and densities. \citet{Chamel08}
  present a computation of $\sigma$ (including impurity scattering)
  for an accreted crust model \citep{Haensel90a}, finding $23 \le
  \log_{10}(\sigma/\mathrm{s}^{-1}) \le 27.4$ for $T=10^7$ K and $10^9 \le
  \rho\mathrm{[g cm}^{-3}\mathrm{]} \le 10^{13}$ (note that this
  density range covers the whole outer crust including neutron
  drip). \citet{Cumming04} find similar values for an accreted crust,
  viz. $\sigma_\mathrm{p}=1.8\times10^{25}\;\mathrm{s}^{-1}(\rho_{14}^{7/6}/T_8^2)$
  for electron-phonon and
  $\sigma_\mathrm{Q}=4.4\times10^{25}\;\mathrm{s}^{-1}\rho_{14}^{1/3}$
  for impurity scattering (provided $Q=1$).}

\cchange{To model a realistic star, we choose the electrical resistivity
  to be $1.3 \times 10^{-27}\,\mathrm{s} \le  \eta_\mathrm{r} \le 1.3 \times
    10^{-24}$ s, covering the range quoted in the previous paragraph. Throughout this paper, we also run 
  simulations with artificially high values of $\eta$, in the range $1.3
  \times 10^{-27} \le (\eta/1\,\mathrm{s}) \le 9.2 \times 10^{-11}$, in order
  to accelerate resistive processes and observe their evolution over a
  computationally practical time interval.}

\change{For simplicity, we assume an isothermal equation of
  state throughout this article. During the late stages of accretion
  ($M_\mathrm{a} \ga 10^{-3}   M_\odot$), however, the magnetic mountain mass is comparable to the mass of
  the neutron star crust and the model mountain contains a wide range of
  densities and temperatures as a function of depth. Pycnonuclear
  reactions in the deep regions 
  ($\rho \ga 10^{12}$ g cm$^{-3}$) feed thermal energy into an
  adiabatic mountain. The assumption of
  isothermality breaks down and a realistic equation of state for
  non-catalyzed matter is required \citep{Haensel90a}. In particular,
  the accreted material is expected to solidify at densities $\ga 10^8$ g
  cm$^{-3}$ \citep{Haensel90b} and will sink into the crust, which
  needs to be modelled as an elastic solid \citep{Ushomirsky00}. In a
  self-consistent model, the electrical conductivity will be computed
  as a function of $\rho$ and $T$. Furthermore, a strong magnetic
  field ($B \gg 10^9$ G) breaks the symmetry of 
  electron transport processes and causes an anisotropic
  conductivity \citep{Potekhin99b}. The effect of
  a realistic equation of state is subject of current work and the results
  will be presented elsewhere.}

\change{In light of the discussion above, it is not immediately
  obvious at what location in the crust $\sigma$ needs to be
  evaluated. In the end, however, we note that there are other
  deficiencies in our model which outweigh the uncertainties in the
  conductivity (most notably, sinking). In the context of this article,
  we therefore treat $\sigma$ as a fiducial parameter. In particular,
  we will show how the resistive relaxation time scales with $\sigma$
  in \sref{relaxation_time}.}




\section{Resistive instabilities}
\label{sec:resistive_instabilities}
\begin{table}
  \centering
  \caption{Simulation parameters. $\eta$
  measures the resistivity in terms of the realistic value
  $\eta_\mathrm{r}=1.3 \times 10^{-27}$ s and also determines the Lundquist number
  $Lu=\tau_\mathrm{D}/\tau_\mathrm{A}$, i.e. the ratio of the
  resistive time-scale $\tau_\mathrm{D}$ to the Alfv\'{e}n time-scale
  $\tau_\mathrm{A}$. Models A--D are axisymmetric; models E--H are
  nonaxisymmetric. All models are for $M_a=M_c$.}
  \begin{tabular}{@{}ccccp{0.1\textwidth}}
    \hline
    Model & $\log_{10} (\eta/\eta_\mathrm{r})$ & $Lu$ & axisymmetric \\
    \hline
    A &  1 & $5.7\times 10^{15}$ & yes  \\
    B &  14.9 & $8.0 \times 10^0$ & yes  \\
    C &  15.9 & $8.0 \times 10^{-1}$ & yes  \\
    D &  16.9 & $8.0 \times 10^{-3}$ & yes  \\
    \hline
    E &  1 & $2.99 \times 10^{14}$ & no \\
    F &  14.9 & $4.25 \times 10^3$ & no \\
    G &  15.9 & $4.25 \times 10^2$ & no \\
    H &  16.9 & $4.25 \times 10^1$ & no \\
    \hline
\end{tabular}
  \label{tab:models}
\end{table}

In general, MHD systems with a finite conductivity exhibit a
plethora of resistive instabilities acting on time-scales much shorter
than the diffusion time-scale \citep{Lifschitz89, Biskamp93}. Our first
task is to find out if such instabilities are present
here and on what time-scales they act. Table \ref{tab:models} lists
the simulations performed to this end. We track the evolution of the mass
ellipticity $\epsilon$ as a convenient way to parametrize the
evolution of the global hydromagnetic structure (VM08).

\subsection{Axisymmetric dynamics}
\label{sec:axisym:general}
\fref{axisym:ell_with_time} displays $\epsilon$ as a
function of time for models A--D in table \ref{tab:models}.
In order to find out if a genuine instability grows on an
$e$-folding time-scale
$\tau_\mathrm{I} < \tau_\mathrm{D}$, we artificially increase $\eta$
and hence the Lundquist number $Lu=\tau_\mathrm{D}/\tau_\mathrm{A}$
(models B--D). Here, $\tau_\mathrm{A}=L \rho^{1/2}/B$ and
\change{$\tau_\mathrm{D}=L^2 \sigma$} denote the Alfv\'{e}n and the diffusion
time-scales, respectively, \change{$\sigma$ is the electrical
  conductivity} and  $L=(|\mathbf{B}|/|\nabla^2 \mathbf{B}|)^{1/2}$ is a
characteristic length-scale. Clearly, $L$, $\tau_\mathrm{A}$,
and $\tau_\mathrm{D}$ are functions of position and time. We minimize
$L$ and $\tau_\mathrm{A}$ over the integration volume, finding (for
the axisymmetric models) $L=1.05 \times 10^{-3}
h_0$ and $\tau_\mathrm{A}=105 \tau_0$ respectively.

During the first oscillation cycle in model B ($Lu=8$), $\epsilon$
declines more steeply than in model A before 
tapering off. This behaviour becomes more distinct in model
C ($Lu = 0.8)$, where $\epsilon$ decreases rapidly, then plateaus
when the equatorward motion of the mountain stops and
subsequently reverses. This cycle of decline followed by plateauing
repeats several times while $\epsilon$ tends to zero overall.

Particularly interesting from a physical point of view is the
behaviour of model D, with $\tau_\mathrm{D} \ll \tau_\mathrm{A}$. As $\eta$ is large, the
magnetic field is unable to contain the mountain at the 
magnetic pole. Consequently, the plasma slips through the
field and falls towards the magnetic equator, where it is
reflected at the boundary; that is, the mountain meets its
counterpart centred at the other pole. As a result, $\epsilon$
oscillates around the abscissa. A realistic neutron star
never enters the regime $\tau_\mathrm{D} \ll \tau_\mathrm{A}$, but the
tendency of the mountain to slip and bounce affects the dynamics
for all values of $\tau_\mathrm{D}/\tau_\mathrm{A}$, as discussed in
\sref{instabilites:oscillation}.

\begin{figure}
  \includegraphics[width=84mm, keepaspectratio]{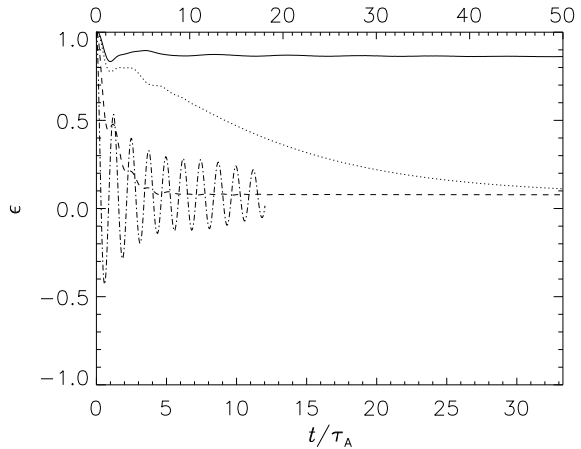}
  \caption{Evolution of mass ellipticity $\epsilon$ for different
    Lundquist numbers (from top to bottom)
    $Lu=5.7 \times 10^{15}, 8, 0.8, 8 \times 10^{-3}$
    (solid, dotted, dashed, dash-dotted) 
    for the axisymmetric models A--D with $M_a=M_c$. The time is
    measured in units of the Alfv\'{e}n time (bottom axis). The top
    axis measures the time in units of the diffusion time for model
    C. Clearly, $\epsilon$ decays on the diffusive time-scale.} 
  \label{fig:axisym:ell_with_time}
\end{figure}

\begin{figure*}
  \includegraphics[width=168mm, keepaspectratio]{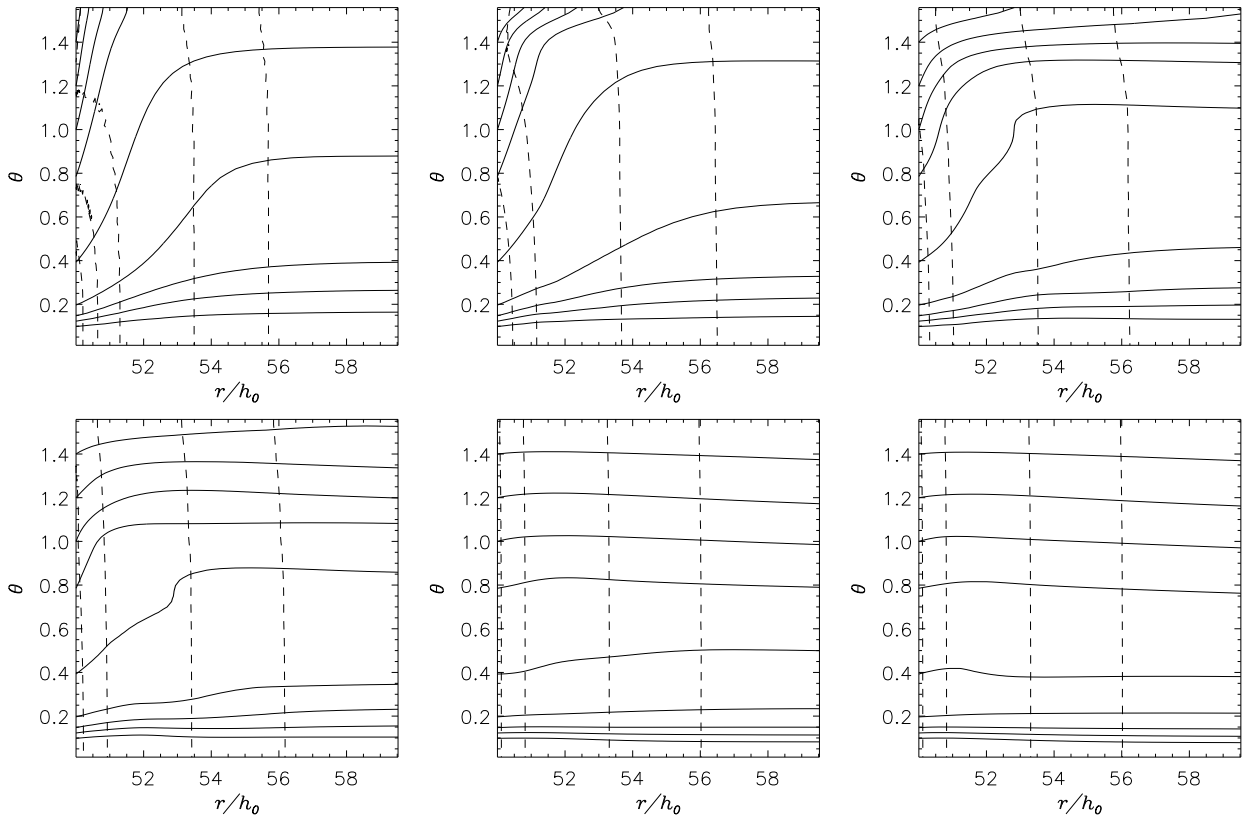}
  \caption{Meridional section of model C at $t/\tau_\mathrm{A}=0, 0.428,
    1.09, 1.76, 4.76, 5.80$ (top left to bottom right). Shown are
    density contours (dashed curves) with values 
    $\log_{10}(\rho/\rho_0')=-13, -12, -11, -10.7, -10.5, -10.3$
    and magnetic flux surfaces in cross-section (solid curves). The plasma
    diffuses through the flux surfaces while the magnetic field
    relaxes radially.}
  \label{fig:axisym:modC_2d}
\end{figure*}

\fref{axisym:modC_2d} shows the density contours (dashed curves) and
projected magnetic flux surfaces (solid curves) for a
meridional slice of model C. Snapshots are taken at $t/\tau_\mathrm{A}=0, 0.428,
    1.09, 1.76, 4.76, 5.80$. At $t/\tau_\mathrm{A}=0.428$ and $1.76$,
$\epsilon$ is in decline, according to \fref{axisym:ell_with_time} (model C, dashed
line). At $t/\tau_\mathrm{A}=0, 1.09$ and $4.76$,
$\epsilon$ is in a plateau. The configuration settles down at $t=5.80
\tau_\mathrm{A}$.

The oscillations in \fref{axisym:ell_with_time} and
\fref{axisym:modC_2d} are driven by the hydrostatic pressure gradient
perpendicular to the magnetic flux surfaces. Their amplitude remains
bounded. Pressure-driven
instabilities, such as the interchange or ballooning mode, 
grow when the field line curvature has a
component along the pressure gradient (i.e. $\bkappa \bcdot \nabla p > 0$,
where $\bkappa = \bmath{b} \bcdot \nabla \bmath{b}$ and
$\bmath{b}=\bmath{B}/B$), a configuration termed unfavourable
curvature \citep{Lifschitz89}. The top left panel of
\fref{axisym:modC_2d} shows clearly that the pressure gradient (which
is proportional to the density gradient) in the
ideal-MHD equilibrium is opposed to the curvature,
preventing the onset of a pressure-driven instability.
Line tying also contributes to stability (VM08).

\label{sec:axisym:general}

\subsection{Nonaxisymmetric dynamics}
\begin{figure}
  \includegraphics[width=84mm, keepaspectratio]{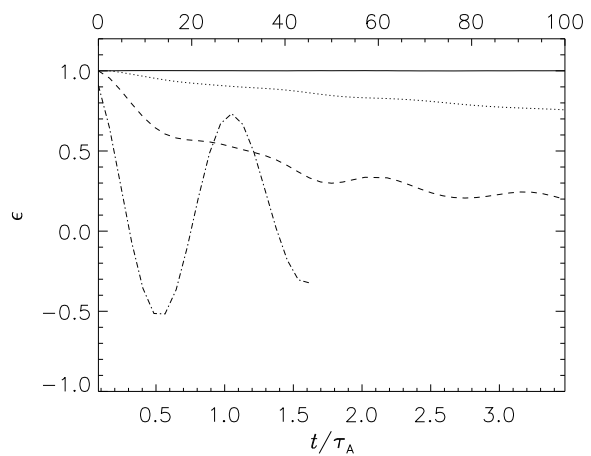}
  \caption{Evolution of mass ellipticity $\epsilon$ for different
    Lundquist numbers (from top to bottom) $Lu=2.99 \times 10^{14}, 4.25 \times
    10^3, 4.25 \times 10^2, 42.5$ (solid, 
    dotted, dashed, dash-dotted) for the nonaxisymmetric models
    E--H with $M_a=M_c$. The time is measured in units of the
    Alfv\'{e}n time (bottom axis). The top axis measures the time in
    units of the diffusion time for model G. As in \fref{axisym:ell_with_time},
    $\epsilon$ decays on the diffusive time-scale.}
  \label{fig:axisym:ell_with_time_3d}
\end{figure}

The stability of an MHD system changes considerably upon passing from
two to three dimensions. It turns out that, in the ideal case,
the additional degree of freedom accomodates toroidal Parker
modes that rearrange the
axisymmetric equilibrium into a slightly nonaxisymmetric state
(VM08). The stability of this state when resistivity is switched on is
the concern of this section. The relevant models are labelled E--H in
table \ref{tab:models}. 

Following \sref{axisym:general}, we first examine the time evolution of
$\epsilon$ for models E--H. The results are summarized in
\fref{axisym:ell_with_time_3d}. Strictly speaking, the 
definition of $\epsilon$ is only meaningful for an axisymmetric
configuration. However, the three-dimensional equilibrium deviates
from axisymmetry by less than 0.8 per cent (VM08), so $\epsilon$ is a
good proxy for the global hydromagnetic structure. We find that
Model E is stable for $t \le 3.5 \tau_\mathrm{A}$. In models F and G,
which have $Lu \la 4.25 \times 10^3$ and $\tau_\mathrm{A}= 124
\tau_0$, the mountain dissipates on the diffusive time-scale
(e.g. $\tau_\mathrm{D}=5.26 \tau_0$ for model G). Model H  ($Lu=42.5$)
exhibits the pressure-driven oscillations observed in model D
(cf. \fref{axisym:ell_with_time}).

\begin{figure*}
  \includegraphics[width=168mm, keepaspectratio]{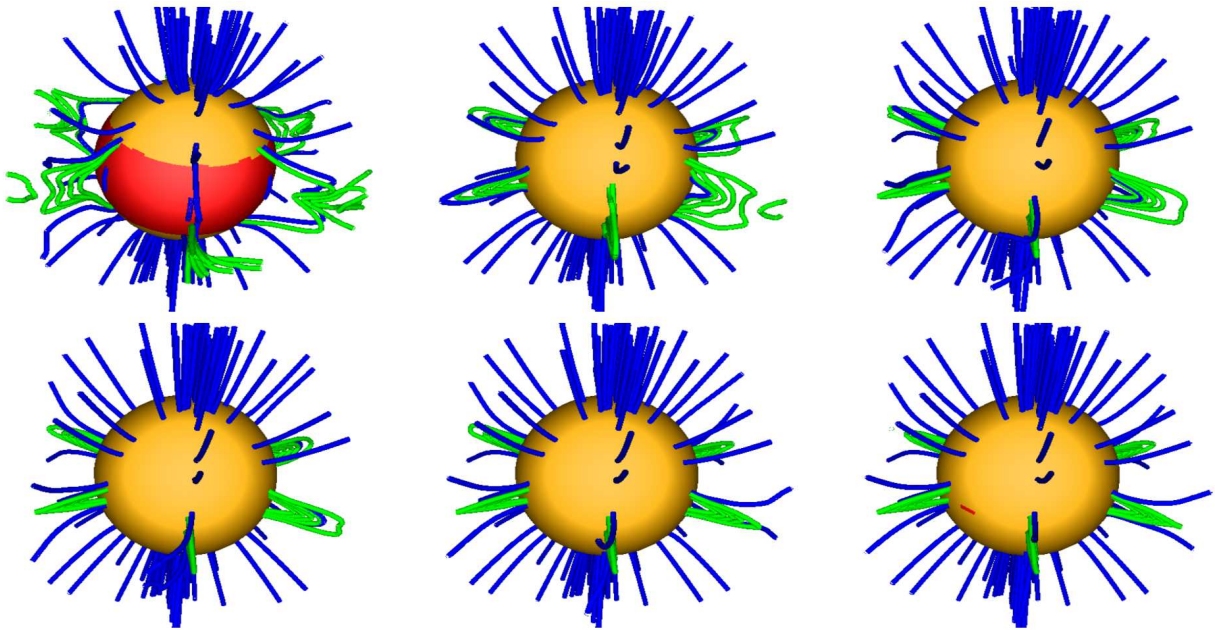}
  \caption{Density and magnetic field structure of model G at
    \change{$\tau/\tau_\mathrm{A}= 0, 1.70, 3.40, 5.09, 6.79, 8.57$} (from top
    left to bottom right). The mountain (orange surface) is  defined by the isosurface
    $\rho(r,\theta,\phi)=1.03\times10^{9}$ g cm$^{-3}$. In order to
    assist with visualization, all length-scales of the mountain and the
    field lines are magnified five-fold. The footpoints of the blue
    field lines start from the stellar
    surface, while green field lines start from the equatorial plane.} 
  \label{fig:stab:modG_3d}
\end{figure*}

The three-dimensional hydromagnetic structure of model G is captured
in a series of snapshots in \fref{stab:modG_3d}. Shown is the mountain
(orange surface), delineated by the isosurface $\rho=1.03 \times 10^9$ g
cm$^{-3}$, along with the magnetic field lines (blue and green curves), at
the instants $t/\tau_\mathrm{A}=0, 1.70, 3.40, 5.09, 6.79, 8.57$.
The initial configuration (top-left panel) is the outcome of the
three-dimensional undulating submode of the Parker instability
(VM08). The field lines curve towards the
magnetic poles, while the orange isosurface spreads equatorwards by
32 per cent relative to its initial position. Soon after
the resistivity is switched on (top-middle panel), the system 
behaves like model C: magnetic tension straightens the field
lines radially, while the plasma slips laterally through
the flux surfaces, allowing the
magnetic mountain to escape its polar confinement and
spread over the neutron star surface. However, the nonaxisymmetric
configuration is the saturation state of the transient Parker
instability. Hence, unlike model C, the global hydromagnetic
oscillations in model G have already died away. The
instability time-scale is given by the diffusion time-scale, not
the tearing-mode time-scale $(\tau_\mathrm{D} \tau_\mathrm{A})^{1/2}$
\citep{Furth63}.

\subsection{Oscillation enhanced diffusion}
\label{sec:instabilites:oscillation}
\begin{figure}
  \includegraphics[width=84mm, keepaspectratio]{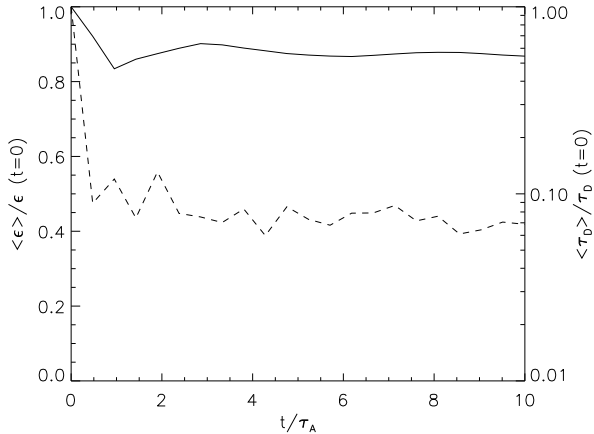}
  \caption{Mass ellipticity $\epsilon$ (solid curve, left linear axis) and
    diffusion time-scale $\tau_\mathrm{D}$ (dashed curve, right logarithmic
    axis) as functions of time, in units of the Alfv\'{e}n time, for
    model A. Both quantities are normalized to their
    initial values. During the first cycle, $\tau_\mathrm{D}$ drops by
    91 per cent.} 
  \label{fig:axisym:oscillation_diffusion_timescale}
\end{figure}
As the axisymmetric mountain oscillates laterally, the field
gradients steepen whenever the field compresses. This effect accelerates
resistive relaxation. \fref{axisym:oscillation_diffusion_timescale} plots
$\tau_\mathrm{D}$ (right, logarithmic axis) and $\epsilon$ (left,
linear axis) as functions of time for model A. During the first cycle,
$\tau_\mathrm{D}$ drops to nine per cent of its original value and
diffusion proceeds proportionally faster.
The effect of diffusion is two-fold. (i) The plasma slips through
magnetic flux surfaces and moves towards the magnetic
equator. Eventually, as seen in the lower middle panel in
\fref{axisym:modC_2d}, it covers the surface evenly and $\epsilon$
decreases (\fref{axisym:ell_with_time}). (ii) Magnetic tension
  causes the field lines to straighten radially. Close to the
magnetic equator, the hydrostatic pressure from the drained plasma
also drives the magnetic field outwards.

\citet{Mouschovias74} showed that an isothermal gravitating MHD
system possesses a total energy $W$, which can be written as the sum of
gravitational ($W_\mathrm{g}$), kinetic ($W_\mathrm{k}$), magnetic
($W_\mathrm{m}$), and acoustic ($W_\mathrm{a}$) contributions, defined
by Eqs. (10)--(13) in VM08.
In ideal MHD, $W$ is a conserved quantity. Adding resistivity
allows the magnetic flux to dissipate, converting $W_m$ to
$W_a$ via a source term $(\gamma-1) \eta |\bmath{j}|^2$ in the energy
equation, where $\gamma$ is the adiabatic index. In an isothermal setup, this
source term vanishes and the energy equation is
trivially satisfied; heat is absorbed by a reservoir.

\begin{figure}
  \includegraphics[width=84mm, keepaspectratio]{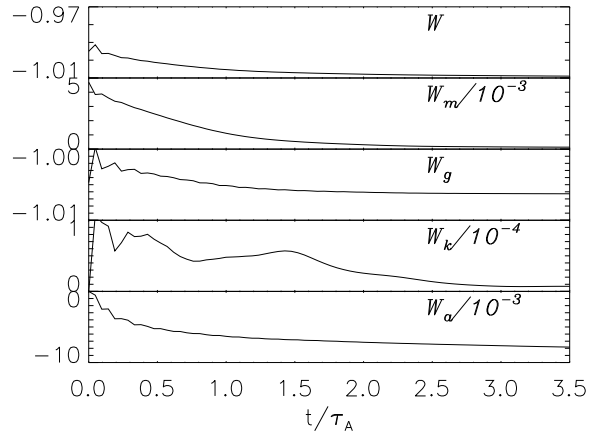}
  \caption{Temporal evolution of the total, magnetic, gravitational, kinetic,
    and acoustic energies $W$, $W_m$, $W_g$,  $W_k$,
    and $W_a$ (top to bottom) for model C, all
    normalised to $W_0=2.4\times10^{36}$ erg and corrected for
    mass loss through the outer border. Note the decrease in $W_m$ due
    to magnetic dissipation.}
  \label{fig:axisym:modC_energy}
\end{figure}
The time dependence of the above four contributions to the energy
integral for the axisymmetric model C are shown in
\fref{axisym:modC_energy}. Following VM08, we
correct for mass loss through the $r=R_m$ border by multiplying
$W_\mathrm{g}$, $W_\mathrm{k}$, and $W_\mathrm{a}$ by $M(t=0)/M(t)$,
where $M(t)$ is the total mass in the simulation volume at time t.
Clearly, some energy is converted to heat: $W$
drops by 1.4 per cent during the interval $t \la 3 \tau_\mathrm{A}$,
as the magnetic field dissipates. $W_g$ decreases because the accreted
matter, which is initially confined at the magnetic pole, distributes itself evenly over the
star's surface. $W_\mathrm{k}$ rises sharply when the whole system
reconfigures and then slowly decreases due to numerical
dissipation. $W_\mathrm{A}$ decreases along with $|\nabla
p|$. 

\begin{figure}
  \includegraphics[width=84mm, keepaspectratio]{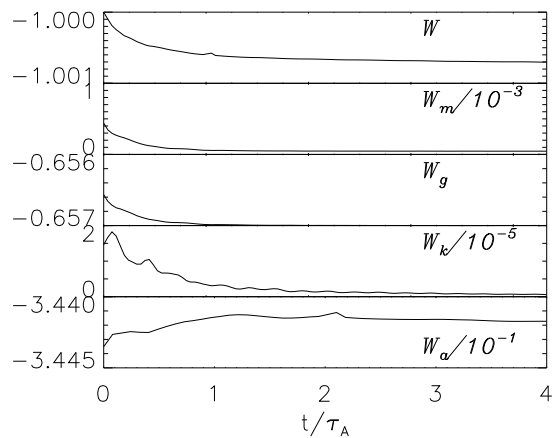}
  \caption{The evolution of total, magnetic, gravitational, kinetic,
    and acoustic energies $W$, $W_m$, $W_g$,  $W_k$, 
    and $W_a$ (top to bottom) for model G, all
    normalised to $W_0=2.1\times10^{36}$ erg, as a function of time
    (in units of the Alfv\'{e}n time) and corrected for mass
    loss through the outer border.}
  \label{fig:nonaxisym:modG_energy}
\end{figure}

\fref{nonaxisym:modG_energy} shows the time dependence of the
different energy contributions for the nonaxisymmetric model
G. Similar to \fref{axisym:modC_energy}, $W$ drops by $\approx 2$ per
cent on the diffusion time-scale. The main losses occur in $W_m$,
which drops by \change{one order of magnitude}, and $W_g$, which decreases by 1 per cent
(from a high base). The kinetic energy slowly rises, as an overstable
mode grows (see \sref{nonaxisym:moments}). Since \textsc{zeus-mp} does
not explicitly include viscosity, $W_k$ dissipates numerically
(i.e. through the grid viscosity).

\change{On the other hand, discretizing the continuous MHD equations
introduces numerical errors that dissipate magnetic energy and can
damp the growth of unstable modes. This numerical viscosity, $\nu$,
can therefore artificially stabilize our configuration. A good measure
for the relative importance of $\nu$ is 
the magnetic Prandtl number, $Pr_m =
\tau_\mathrm{D}/\tau_\mathrm{visc}$, where $\tau_\mathrm{visc}=\rho
l_v^2/\nu$ with $l_v$ being a characteristic length scale for velocity
gradients and $\nu$ the viscosity. Since $\nu$ 
owes its existence to the discretization of the MHD equations, it
depends on the grid size and the field gradients. In order to obtain an accurate
estimate for $Pr_m$, we compute the timescale, $\tau_\mathrm{visc}$, on
which ideal-MHD oscillations of an axisymmetric configuration die away
\citep{Payne06a} and compare it to the diffusive timescale,
$\tau_\mathrm{D}$, finding $Pr_m=1.03 Lu$. Hence, the
contribution of numerical viscosity is generally small (e.g. $Pr_m\approx
10^{-3} \ll 1$ in model D).}





\section{Magnetic field structure}
\label{sec:field_structure}
The global hydromagnetic evolution observed in
\sref{resistive_instabilities}  occurs on the ohmic
time-scale. This indicates that relaxation is dictated by magnetic diffusion
rather than resistive transient instabilities on short time-scales,
such as the large-scale tearing mode or the localized gravitational
mode \citep{Furth63}. Transient instabilites occur in the neighbourhood of
current sheets, which dissolve into magnetic islands and 
dissipate. In this section, we examine the magnetic geometry of the
resistively relaxing mountain to check whether it is consistent with
the above view that diffusion on large scales dominates the
evolution. 

\subsection{Neutral surfaces}
\label{sec:nonaxisym:general}
We begin by investigating the magnetic field structure
of the axisymmetric model C (\fref{axisym:modC_2d}).
The initial equilibrium configuration is depicted in the top-left
panel. Notice that there are no magnetic neutral points present. The
mountain is held in place by the tension of the line-tied magnetic
field. 

\begin{figure*}
  \includegraphics[width=168mm, keepaspectratio]{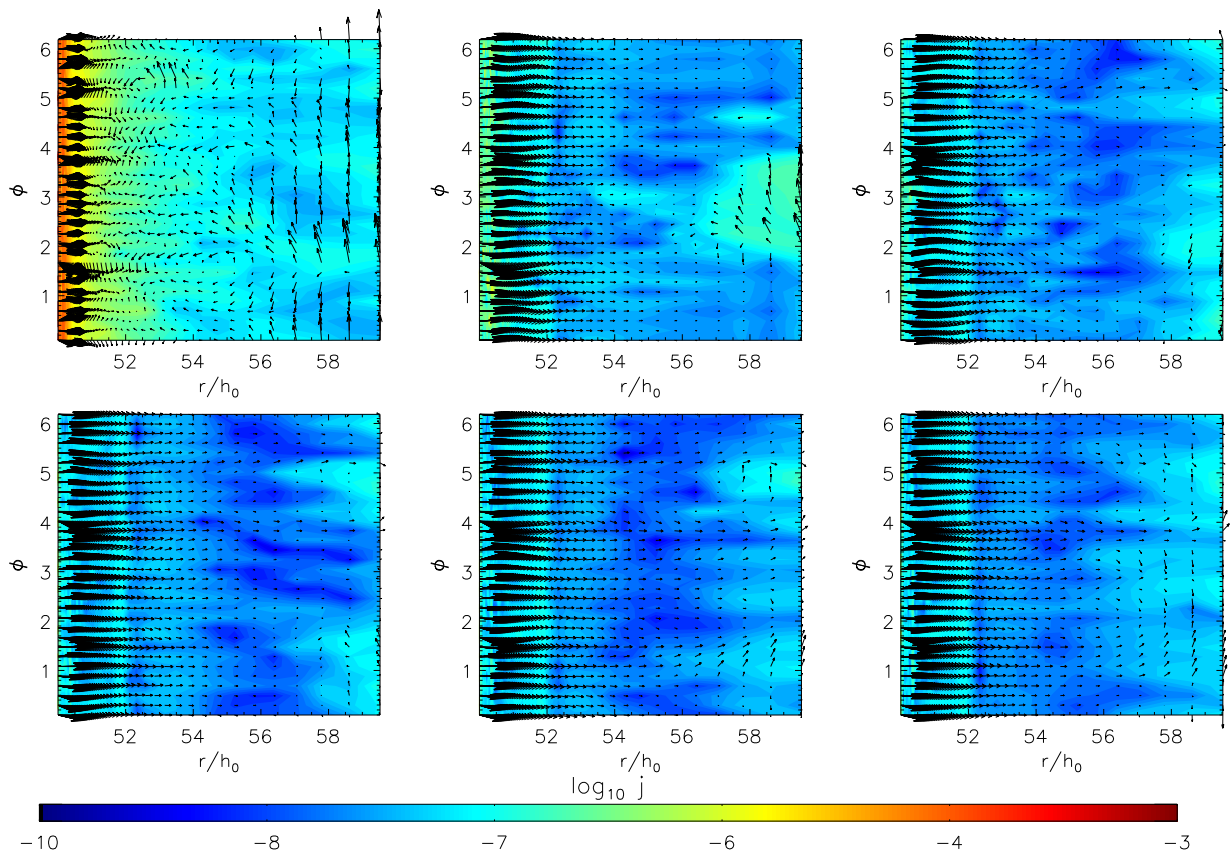}
  \caption{Equatorial slices of the magnetic structure and current
    flows in model G at at $t/\tau_\mathrm{A}=0, 1.69, 3.39, 5.08, 6.77, 8.55$ (from top left to
    bottom right). Shown are the projections of 
    the magnetic field vectors onto the  plane $\theta=1.5$ rad (arrows) and
    the modulus of the current density $|j|$ (color coded).} 
  \label{fig:stab:modJ_2d_equator}
\end{figure*}

Does the configuration contain current sheets? \citet{Hanasz02} showed
that the undulating submode of the Parker instability in a Cartesian
geometry creates current sheets in the plane perpendicular to the
magnetic flux surfaces between regions with alternating
polarity. \fref{stab:modJ_2d_equator} displays a time series of equatorial slices of the
magnetic field from model G at $t/\tau_\mathrm{A}= 0, 1.69, 3.39,
5.08, 6.77, 8.55$. The projection of $\bmath{B}/B$ onto the equatorial  
plane is indicated by arrows, while the current density $|\bmath{j}|$
is color coded. The top left panel shows the initial configuration for
our experiment, generated from an axisymmetric mountain after
the undulating submode of the Parker instability saturates. While
$|\bmath{j}|$ is greatest close to the 
stellar surface, where $B$ is high, long radial current filaments are
also clearly present, albeit not as distinctly as in
\citet{Hanasz02}. The filaments are neutral sheets.

\subsection{Reconnection}
Reconnection occurs at the current sheets in
\fref{stab:modJ_2d_equator}, quickly smoothing the toroidal 
gradients. Line tying at the stellar surface forces the field lines to
adjust into a dipolar configuration. A finite resistivity
therefore acts to restore axisymmetry. In addition, the line-tying
boundary condition acts as a source of magnetic flux, which is thence
transported radially outward by diffusion.

\begin{figure}
  \includegraphics[width=84mm, keepaspectratio]{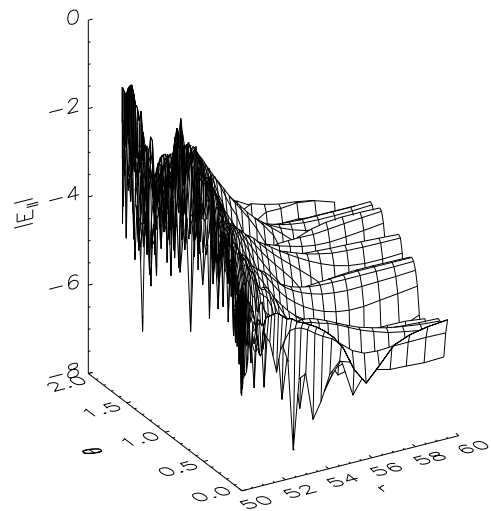}
  \caption{Meridional slice of the local field-aligned electric field
    \change{$\log_{10} |\bmath{E} \bcdot \bmath{B}|$}, normalised to
      \change{$B_0 E_0=1.51 \times 10^{-11} B_0 h_0^{-1}$}
    for model G. Reconnection mainly occurs 
    near the stellar surface in the magnetic belt but also in the outer
    equatorial region (at $r \ga 56 h_0$ and $0.7 \la \theta \la 1.5$).}
  \label{fig:axisym:modelK_parallel}
\end{figure}

Where does reconnection occur? \citet{Schindler88} pointed out that a
necessary and sufficient condition for global magnetic reconnection
along some field line $C$ is that the electric field has a component parallel to $\bmath{B}$,
\begin{equation}
  \int_C \mathrm{d}s\;\bmath{E} \bcdot \bmath{B} \ne 0,
\end{equation}
where the integral is taken along $C$. (Equivalently, the helicity changes with time.)
We plot a meridional slice of $\bmath{E} \bcdot \bmath{B}$ at
$\phi=2.3$ rad in \fref{axisym:modelK_parallel}. Not surprisingly,
\change{$\bmath{E} \bcdot \bmath{B}$} is highest in the magnetic belt region,
close to the star's surface. However,
the undulating submode of the Parker instability also induces
small toroidal currents (top left panel in
\fref{stab:modJ_2d_equator}), so that \change{$\bmath{E} \bcdot \bmath{B}$} is high in
the equatorial region too. We 
integrate \change{$\bmath{E} \bcdot \bmath{B}$} along two sample field lines with
footpoints at $(\tilde{x}_0, \theta_0, \phi_0)=(0, 0.2, 4.03)$ (field
line \flone) and $(0, 1.0, 4.03)$ (field
line \fltwo) and find $\int_C \change{\bmath{E} \bcdot \bmath{B}} =-8.6 \times
10^{-14} B_0$ (field line \flone) and $\int_C \change{\bmath{E} \bcdot
  \bmath{B}} =-3.2 \times
10^{-16} B_0$ (field line \fltwo) respectively. The topology of the
magnetic field is discussed in \sref{structure:topology}, where we 
show that field line \flone\ undergoes reconnection while field line
\fltwo\ does not.


\subsection{Topology}
\label{sec:structure:topology}
\begin{figure*}
  \includegraphics[width=168mm, keepaspectratio]{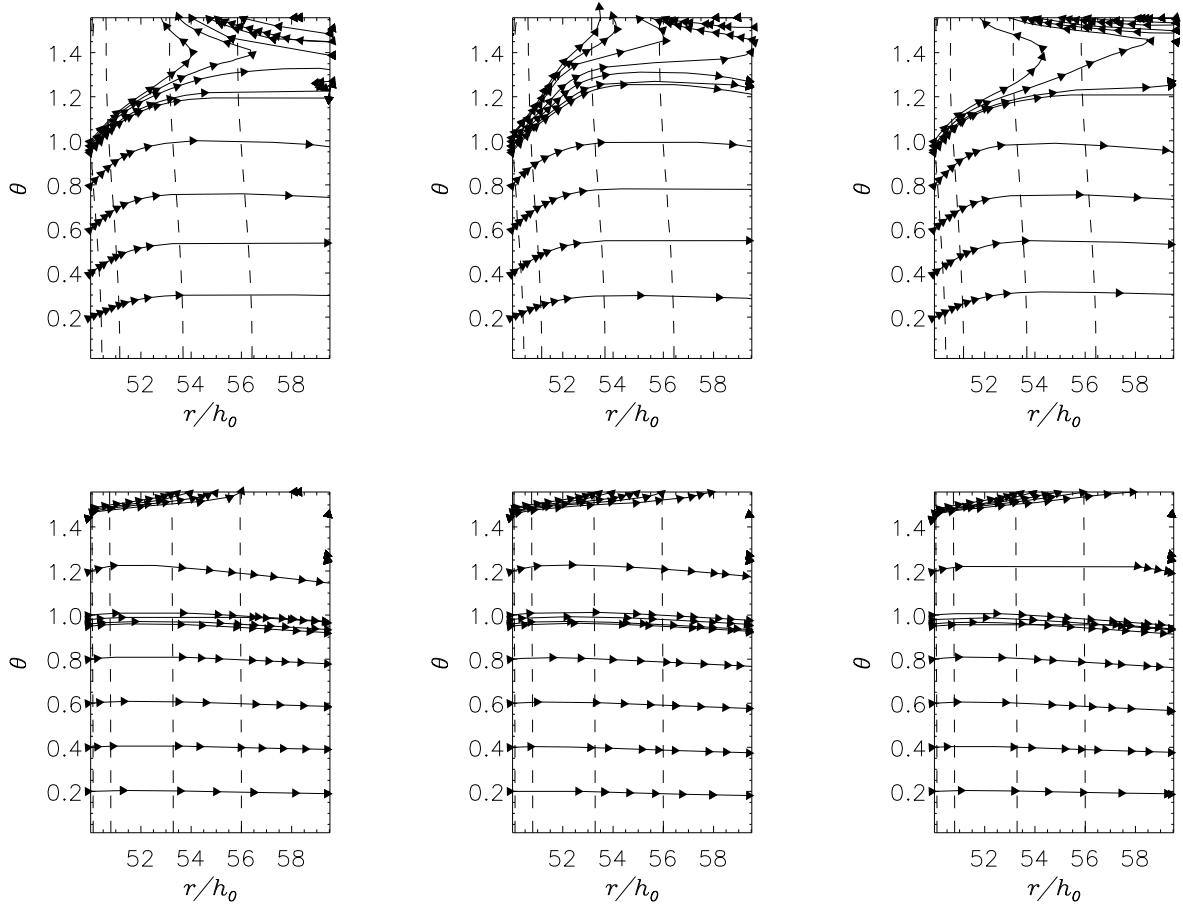}
  \caption{Density contours (dashed) and magnetic field lines (solid)
    for meridional slices at $\phi=0, 1.96, 3.93$ rad (left to right columns)
    and $t/\tau_\mathrm{D}=0, 201$ (top and bottom row) for model
    G. The direction of the magnetic field is indicated by arrows. At
    the Y-point in the top-right corner of each panel, reconnection occurs.} 
  \label{fig:top:field}
\end{figure*}

In this subsection, we briefly discuss the change in magnetic
topology brought about by reconnection. \fref{top:field}
displays the magnetic field lines (solid curves) in three meridional slices
$\phi=0, 1.96, 3.93$ (left, middle, right columns) for model G. One
immediately notices that there is a Y-point
located at \change{$(r, \theta) \approx (56 h_0, 1.4\, \mathrm{rad})$} in
the top-left panel of the figure. The Y-point owes its existence to a boundary
effect in the ideal-MHD simulation: during the onset of the Parker
instability, the plasma is pushed out of the integration volume
through the outer boundary. The subsequent backflow topologically
separates the previously connected field lines.

Associated with the
Y-point is a current sheet at $\theta \approx 1.4$ rad, which meanders
like a band in the $\phi$ direction. 
A current sheet naturally triggers reconnection. The bottom row of
\fref{top:field} shows the same slices as the top row after $0.6
\tau_\mathrm{D}$. Indeed, the field lines have reconnected: they are
not topologically separated anymore, and the current sheet has
vanished. \change{Alternatively, it is conceivable that the current
  sheet moves along with the plasma flow from its initial position at
  $\theta=1.4$ to the upper boundary at $\theta=\pi/2$.}

\begin{figure*}
  \includegraphics[width=80mm, keepaspectratio]{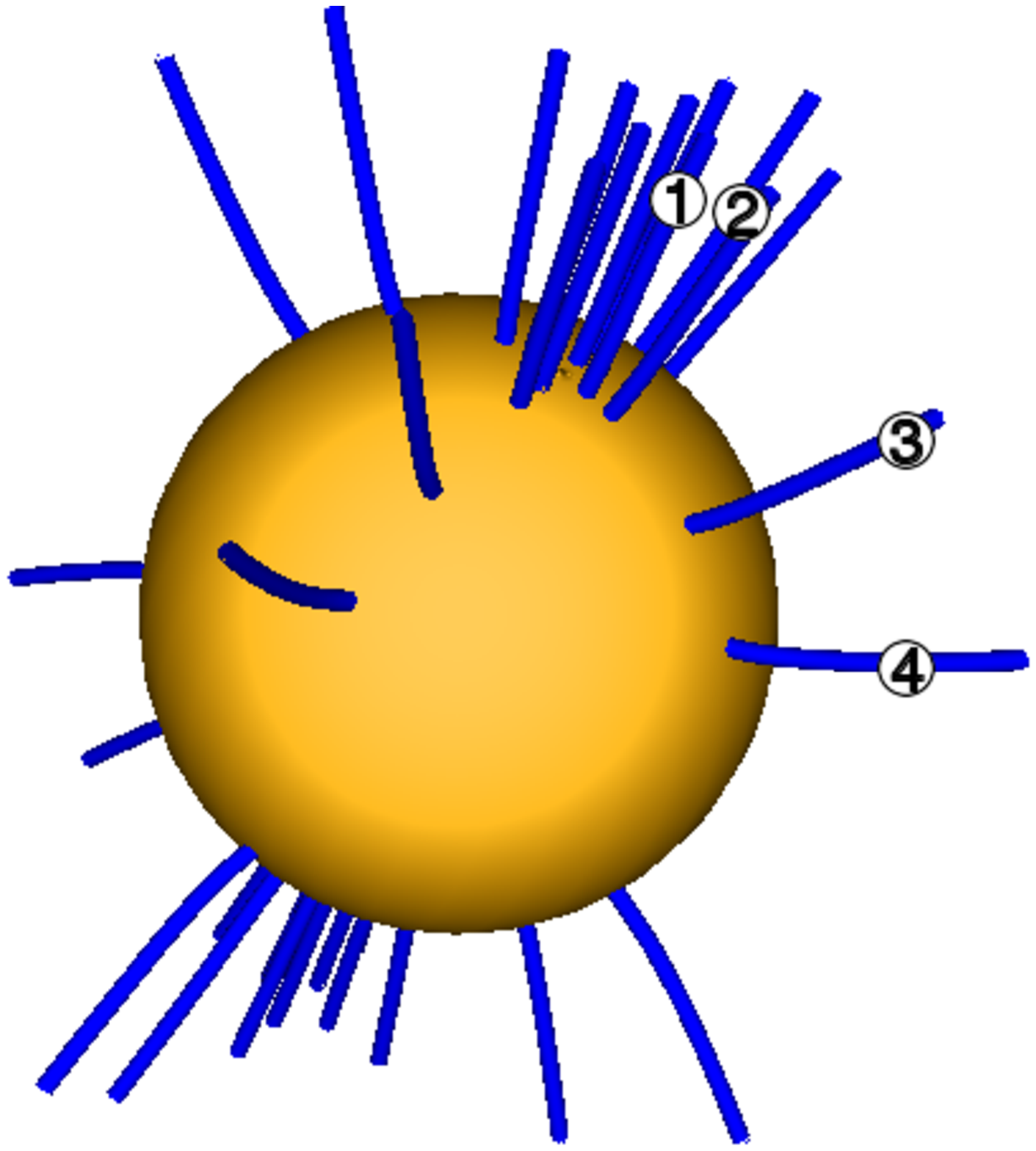}
  \includegraphics[width=80mm, keepaspectratio]{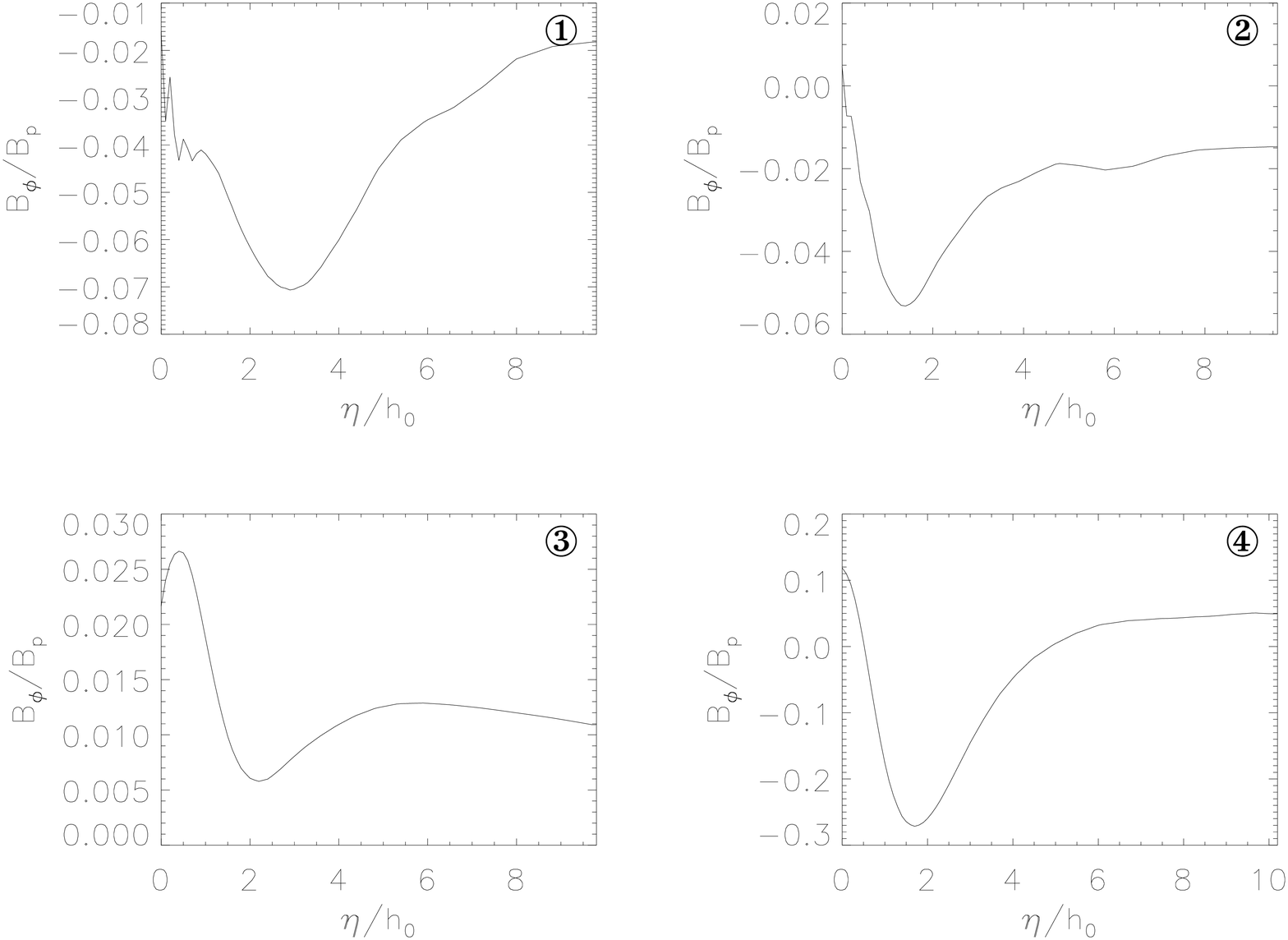}
  \caption{Magnetic pitch angle $B_\phi/B_p$ (right panel) as a function of
    arc length $\eta$ along four magnetic field lines \flone--\flfour\ 
    for model G, for a snapshot taken at $t=0.6
    \tau_\mathrm{D}$. Field lines (blue curves) are identified in the left panel. The 
    mountain is  defined by the orange isosurface
    $\rho(r,\theta,\phi)=1.04\times10^{9}$ g cm$^3$. Red
      indicates the neutron star surface $r=R_\ast$. In order to
    help visualize the structure, all length scales of the mountain and the
    field lines are magnified five-fold.}
  \label{fig:top:pitch_angle}
\end{figure*}
The concept of rational magnetic surfaces,
where the field lines close upon themselves, plays an important role
in a local plasma stability analysis \citep{Lifschitz89}. The bending
of field lines as a result of a Lagrangian displacement $\bxi$ is associated
with an increase in  potential energy, given by $\mathbf{B} \bcdot \nabla
\bxi$. In a tokamak geometry, it can be shown that
this term vanishes on a rational surface, which is directly related to the 
pitch angle $B_\phi/B_p$: the safety factor is defined as
$q=\mathrm{d} B_\phi/\mathrm{d}B_p$, and a rational surface is one where $q$ is a rational number.
In \fref{top:pitch_angle}, we plot the pitch angle as a function of
the arc-length coordinate $\eta$ (right panels) for four
different field lines (labelled \flone--\flfour\ in the left panel) in
model G. Close to the pole  (lines \flone--\flthree), the pitch angle stays below $\approx 3$ per
cent. For line \flfour, it increases towards the equator, ultimately
reaching $\approx 20$ per cent. The zero crossing for \flfour\
indicates that $B_\phi$ changes sign, a relic of the undulating
submode of the Parker instability which gives birth to this
state. Diffusion does not eliminate the toroidal component completely.

We attempted for completeness to characterize the magnetic topology
near the neutral surface using scale invariants of the strain tensor
$\partial B_i/\partial x_j$ \citep{Chong90, Parnell96, Peralta08}, but
this approach yields ambiguous results in this instance.
 
\section{Relaxation time}
\label{sec:relaxation_time}
We are now in a position to compute how long it takes for a
magnetically confined mountain to relax resistively, given $\eta$ and
$M_a$. Ultimately, as $t \rightarrow \infty$, the mountain spreads
itself uniformly over the stellar surface [i.e. $\rho=\rho(r)$],
threaded by a dipole field (i.e. $j=0$ everywhere). However, this
process does not approach completion for realistic $\eta$ over the
lifetime of an accreting neutron star.

Let us define the ohmic relaxation time to be the time that elapses
before the mountain relaxes to $\mathrm{e}^{-1}$ its initial
ellipticity. \fref{axisym:ell_with_time} presents $\epsilon(t)$ for an
axisymmetric mountain with $M_a=M_c$ as a function of
the conductivity $\sigma=\eta^{-1}$. Reading off $\tau_\mathrm{I}$
from $\epsilon(\tau_\mathrm{I})=\mathrm{e}^{-1} \epsilon(0)$, and
fitting the trend by linear least squares, we obtain
\begin{equation}
  \change{\frac{\tau_\mathrm{I}}{\tau_0} = 1.7 \times 10^{-3} \sigma,}
\end{equation}
where $\sigma$ is measured in units of $\sigma_0=\eta_0^{-1}=1.86
\times 10^6$ s$^{-1}$.
For the upscaled star with a realistic $\eta_\mathrm{r}$ we find
\change{$\tau_\mathrm{I} = 6.3 \times 10^6$ yr}, which is \change{comparable
  to} the fiducial accretion time-scale
$\tau_\mathrm{acc}=10^6-10^7$ yr.

\fref{axisym:ell_with_time_3d} presents $\epsilon(t)$ for a
nonaxisymmetric mountain with $M_a=M_c$. Applying the same procedure
from the previous paragraph to \fref{axisym:ell_with_time_3d}, we find
\begin{equation}
  \change{\frac{\tau_\mathrm{I}}{\tau_0}=0.02 \sigma.}
\end{equation}
For an upscaled neutron star with realistic $\eta_\mathrm{r}$, we find 
\change{$\tau_\mathrm{I}=7.6 \times 10^7$ yr}, comparable to the fiducial
accretion time-scale. Astrophysically,
this is the key result of this paper: magnetic mountains in three
dimensions relax resistively over \cchange{$\sim 10^5-10^8$ yr, 
(depending on the particular value of $\sigma$; see \sref{model:resistivity}), not over shorter
time-scales like $\tau_\mathrm{A}$ and $(\tau_\mathrm{A} \tau_\mathrm{D})^{1/2}$.} Note
that $\tau_\mathrm{D}$ is a local quantity for a stationary mountain;
$\tau_\mathrm{I}$ is a better measure of the global diffusion time.

We compare $\tau_\mathrm{I}$ to the growth time of the resistive Parker
instability, whose dispersion relation is calculated in appendix
\ref{sec:app:parker}. The growth time is shortest for short-wavelength
modes and is therefore set by the grid scale ($k\approx 6.6
h_0^{-1}$) in our units. Also, \change{the ratio of magnetic pressure
  to gas pressure}, $\alpha$, is maximal in
the magnetic belt region, where the magnetic pressure balances the gas
pressure, viz. $\alpha \approx 1$. The growth rate is independent of
$k$ and is given by $\Gamma = [1/2 (1-2 \alpha)^2] [\alpha/(1+\alpha)]
(g^2/k^2 u^2) (\i/\tau_\mathrm{D})$, where
$\tau_\mathrm{D}$ is the time required to diffuse over one scale
height. Applying \eref{parker:growing} to the axisymmetric model C, we
find the Parker growth-rate to be $\Gamma = 4.07 \times 10^{6}
\tau_0^{-1}$. \fref{axisym:ell_with_time} shows clearly that
$\tau_\mathrm{I} \gg \Gamma^{-1}$, further supporting our conclusion that the resistive
relaxation occurs on the diffusion time-scale and does not involve
MHD instabilities. The same conclusion applies for the
nonaxisymmetric model G, with the same growth rate as for the
axisymmetric model.

\begin{figure}
  \includegraphics[width=84mm, keepaspectratio]{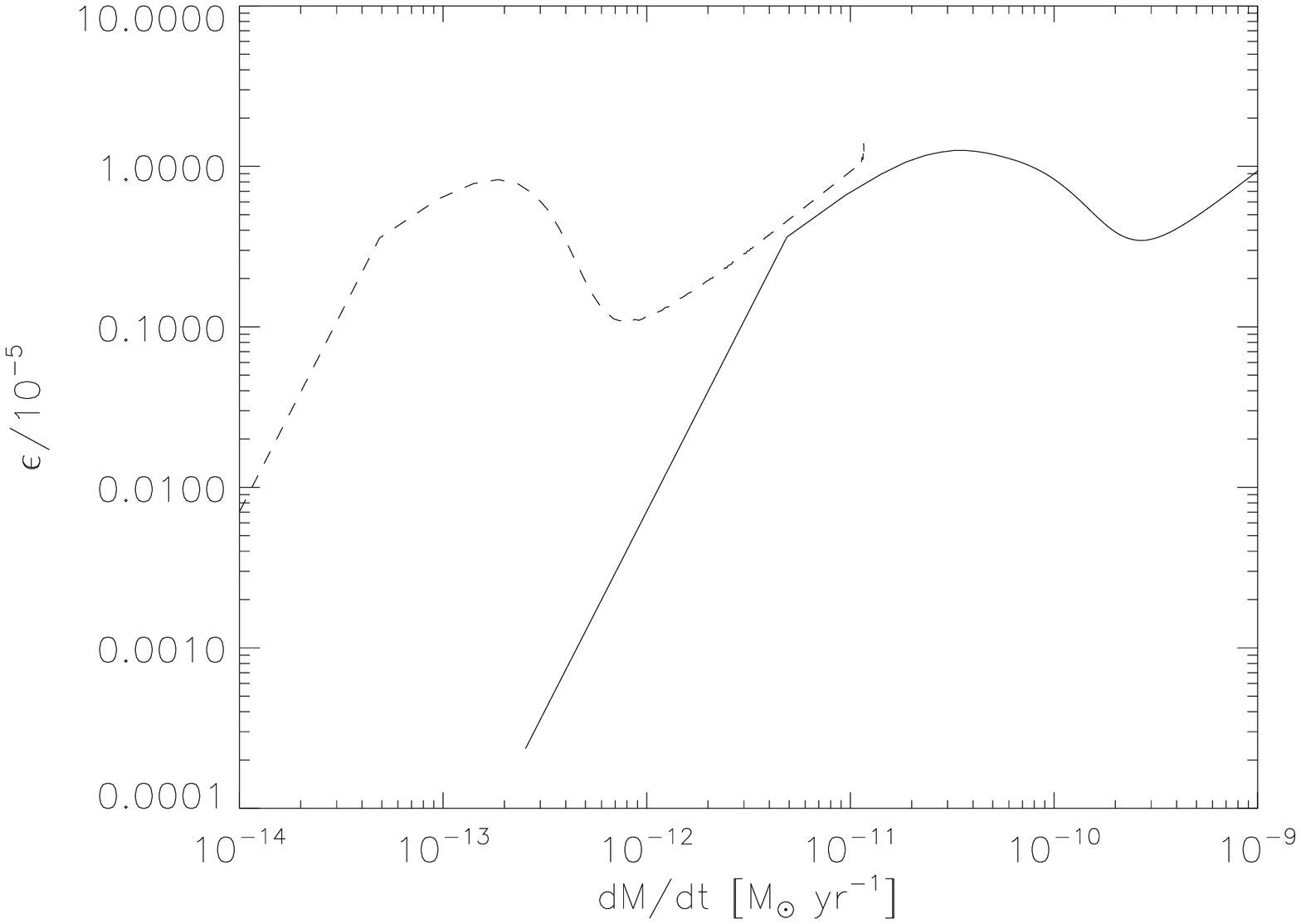}
  \caption{Mass ellipticity $\epsilon$ versus accretion rate $\dot{M}$
    (in units of $10^{-4} M_\odot$ yr$^{-1}$) for an axisymmetric
    grown mountain [see text and \citet{Vigelius08b}]. The solid
    (dashed) curve represents a mountain aged $t=10^6$ yr ($t=10^8$
    yr). Although $\epsilon$ is generally
    higher for the older mountain, since more mass has been accreted,
    it almost touches the curve for the younger mountain, because
    the resistive instability acts to reduce $\epsilon$.}
  \label{fig:grow:epsmadot}
\end{figure}

Another way to present the results on $\tau_\mathrm{I}$ is to ask how
$\epsilon$ varies with the accretion rate $\dot{M}$.  The simulations
underlying \fref{grow:epsmadot} differ from the others in this paper
in one important respect: the mountain is grown  \change{from scratch}
over time \change{(starting from $M_a=0$)}, with mass
injected at the poles of an initially dipolar magnetic field, at a
rate $\dot{M}$ and with $\eta \ne 0$ throughout the experiment. In
other words, resistive relaxation \emph{competes simultaneously} with
accretion. By contrast, in
Figs. \ref{fig:axisym:ell_with_time}--\ref{fig:top:pitch_angle}, a
Grad-Shafranov equilibrium is imported into \textsc{zeus-mp}, $\eta$
is switched on at $t=0$, and the mountain subsequently
relaxes. Growing the mountain confers several advantages: it reflects
the astrophysical process of burial more faithfully and enables us to
reach $M_a = 10 M_c$, cf. $M_a \le 1.4 M_c$ with the Grad-Shafranov
method. The disadvantage is that, at present, we cannot study how the
mountain relaxes after accretion stops, because \textsc{zeus-mp} fails
when the injection ``nozzles'' are turned off (suddenly or with
taper), due to a numerical instability (the grown mountain contains
nonzero flows). We are therefore unable to compare the two numerical
experiments exactly, although they are in close qualitative
agreement.  A detailed explanation of the injection algorithm and 
verification tests can be found in \citet{Vigelius08b}.

\change{A crucial question is whether the Grad-Shafranov equilibria
  can be uniquely attained as accretion onto the magnetic poles
  occurs, in particular, when $\eta \ne 0$. The experiments
  conducted by growing the mountain ab initio [\fref{grow:epsmadot}
  and \citet{Vigelius08b, Vigelius08e}] mimic time-dependent
  accretion more faithfully. The infalling plasma continuously deforms
  an initially dipolar field and every snapshot represents the
  equilibrium configuration for a particular $M_a$. These equilibria
  are in good agreement with previous results obtained analytically or
  numerically with the Grad-Shafranov code [cf. Fig. 4.11 in
  \citet{Vigelius08b}]. In   particular, we find no evidence for
  (ideal or resistive) instabilities occuring in the low-$M_a$
  regime.}

\change{On the other hand, the unavoidably finite size of the simulation box
leads to a subtle uniqueness problem. The material that is added to
the pole pushes the field lines towards the equator. Because of the
boundary conditions we use, these field lines jump discontinuously when
they touch the bottom right-hand corner of the box from $\partial_r
\bmath{B}=0$ when penetrating the boundary $r=R_m$ to $B_r=0$ when penetrating 
the boundary $\theta=\pi/2$ (compare the top-right corner of the
top-right and bottom-left panels in \fref{top:field}). In effect, this is a "reconnection-type" event
which changes the topology of the field lines, their connectivity
to the "outside world", and therefore the effective functional form
of $dM/d\psi$ (which we assume to be constant throughout the run).
In practice, it is likely the effect is very small, the evidence
being (i) the small mass outflow ($\la 1$ per cent of the total mass)
through $r=R_m$ during a typical run, and (ii) the very similar equilibria obtained
from solving the Grad-Shafranov equation and growing the mountain
ab initio \citep{Vigelius08b}. In principle,
though, it can lead to different final states if the mountain is
grown with and without resistivity turned on\footnote{Sterl Phinney,
  private communication}.}

\fref{grow:epsmadot} displays the
ellipticity as a function of $\dot{M}$ for a young ($t=10^6$
yr, solid curve) and an old ($t=10^8$ yr, dashed curve) object.
To perform the simulation over a practical length of time, we artificially
increase $\eta$ to $10^{13.85} \eta_\mathrm{r}$ (solid curve) and
$10^{15.85} \eta_\mathrm{r}$ (dashed curve). We then use the scaling
$\tau_\mathrm{I} \propto \eta^{-1}$ derived from
Figs. \ref{fig:axisym:ell_with_time} and
\ref{fig:axisym:ell_with_time_3d} to relate the results to
astrophysical time-scales. \change{We point out that each curve in
  \fref{grow:epsmadot} basically displays $\epsilon(t)$ and we can
  relabel the abscissa using $M_a = \dot{M} t$.}

There are two opposing effects in the
figure. First, the older object has generally higher $\epsilon$ for a
given $\dot{M}$, simply because $M_a$ is higher. Second, resistive
relaxation has more time to reduce $\epsilon$ in the older object, so
the two curves almost touch at $\dot{M} \approx 5\times 10^{-8}
M_\odot$ yr$^{-1}$. \change{The injection algorithm induces
  global hydromagnetic perturbations; these numerical artifacts are
  visible as oscillations at the high-$\dot{M}$ end of either curve.}

How does our relaxation time compare to previous estimates?
In the small-$M_a$ regime, \citet{Melatos05} found analytically
that resistive relaxation stalls mountain growth at $\epsilon \sim 10^{-5}$
(assuming electron-phonon scattering with a crustal
temperature of $T=10^8$ K). Our results suggest that a mountain with
$M_a=M_c=1.2 \times 10^{-4} M_\odot$ relaxes resistively over \change{$\sim
10^5-10^8$ yr}. Furthermore, when accretion and relaxation proceed
together, we find again that $\epsilon$ saturates at
$\sim 10^{-5}$, even for $M_a > M_c$, in accord with \citet{Melatos05}.

Similar estimates were given by \citet{Brown98} who evaluated the
diffusion time in the crust. Taking into account electron-phonon and
electron-impurity scattering, they found that \change{phonon scattering
dominates impurity scattering (provided $Q \la 1$)} and $\tau_\mathrm{D} \sim
10^4$ yr when the star accretes at the Eddington rate. However, these
authors considered only spherically symmetric accretion and disregarded
the global magnetic structure. \citet{Cumming04} found
$\tau_\mathrm{D} \sim 10^8$ yr for a crustal temperature of $T=10^6$
K, in accord with our results. 

\begin{figure}
  \includegraphics[width=84mm, keepaspectratio]{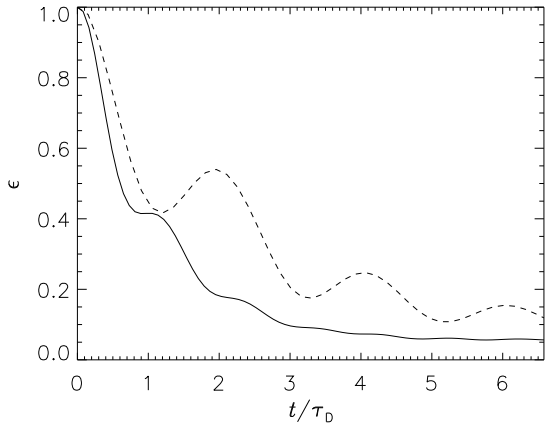}
  \caption{Evolution of mass ellipticity $\epsilon$ as a function of
    time in units of the respective diffusion times
    $\tau_\mathrm{D}$ for axisymmetric models with $M_a = 0.6 M_c$ (solid 
    curve) and $M_a = 1.4 M_c$ (dashed curve). The Lundquist number
    for both models is $Lu=1$. The diffusion time is
    $\tau_\mathrm{D}=123 \tau_0$ ($\tau_\mathrm{D}=66.4 \tau_0$) for
    $M_a=0.6 M_c$ ($M_a=1.4 M_c$).} 
  \label{fig:axisym:ell_massvar}
\end{figure}
We conclude this subsection with a brief discussion of how the
relaxation time changes with $M_a$. \fref{axisym:ell_massvar} compares
$\epsilon$ for two axisymmetric models with different accreted masses,
$M_a=0.6 M_c$ (solid curve) and $M_a=1.4 M_c$ (dashed curve), but the
same Lundquist number $Lu=1$. Note that $\epsilon$ is displayed as a
function of time in units of the respective diffusion time,
$\tau_\mathrm{D}=123 \tau_0$ ($\tau_\mathrm{D}=66.4 \tau_0$) for
$M_a=0.6 M_c$ ($M_a=1.4 M_c$). The magnetic field of the $M_a=1.4 M_c$
model is more distorted and, consequently, $\tau_\mathrm{D}$ is
shorter. Both models exhibit resistive relaxation on the diffusion
time-scale.

\section{Reemergence of the buried magnetic field}
\label{sec:reemergence}
\subsection{Magnetic dipole moment}
An important diagnostic of the global magnetic structure is its magnetic
dipole moment. This integrated value has the advantage
that it is observationally accessible \citep{vanDenHeuvel95}. Indeed,
the observed reduction of the magnetic dipole moment by accretion
is a key motivation of the magnetic mountain concept
(PM04). 

Following VM08, we define the magnetic multipole moment tensor as
\begin{equation}
    d_{ij}(r)=r^{i+1} \int \mathrm{d}\Omega\; Y_{ij}^\ast
   \bmath{r}\bcdot\bmath{B},
\end{equation}
where $Y_{ij}$ denotes the spherical harmonics and $\bmath{r}$ is the
position vector. Henceforth, we evaluate
$d_{ij}$ at the simulation boundary $r=R_m$ and drop $r$.

\begin{figure}
  \includegraphics[width=84mm, keepaspectratio]{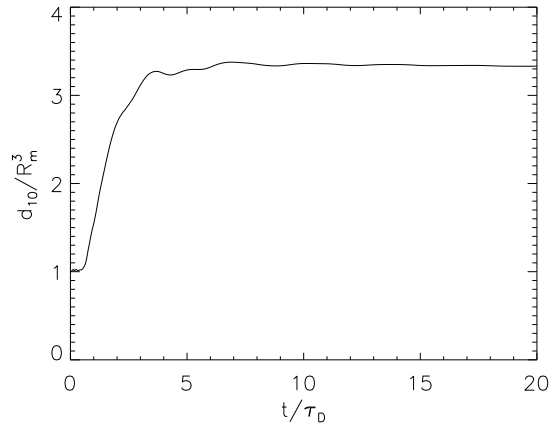}
  \caption{Normalized magnetic dipole moment $d_{10}/R_m^3$ versus
    time (in units of the diffusion time) for
  model C. The resisitive instability relaxes the magnetic field 
  radially, such that $d_{10}$ peaks at $t=7
  \tau_\mathrm{D}$. Eventually, the screening currents in the mountain
  dissipate and $d_{10}$ tends to its initial value.}
  \label{fig:axisym:modC_magmoments}
\end{figure}

\begin{figure}
  \includegraphics[width=84mm, keepaspectratio]{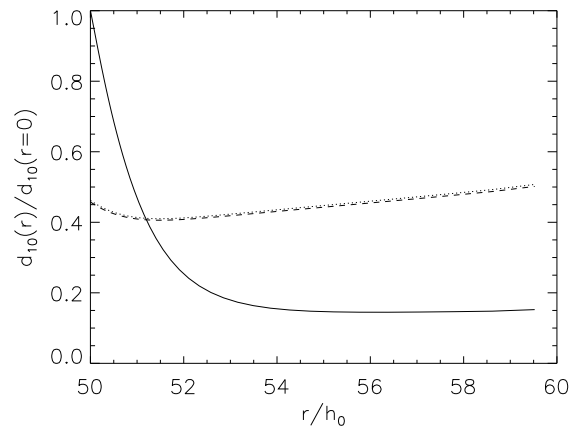}
  \caption{Magnetic dipole moment $d_{10}(r)$ [normalised to
    $d_{10}(r=0)$] versus radius (in units of $h_0$)
    for model C at times $t/\tau_\mathrm{D}=0, 333, 665$ (solid,
    dotted, dashed curves).}
  \label{fig:axisym:modC_rad_dipole}
\end{figure}

\begin{figure}
  \includegraphics[width=84mm, keepaspectratio]{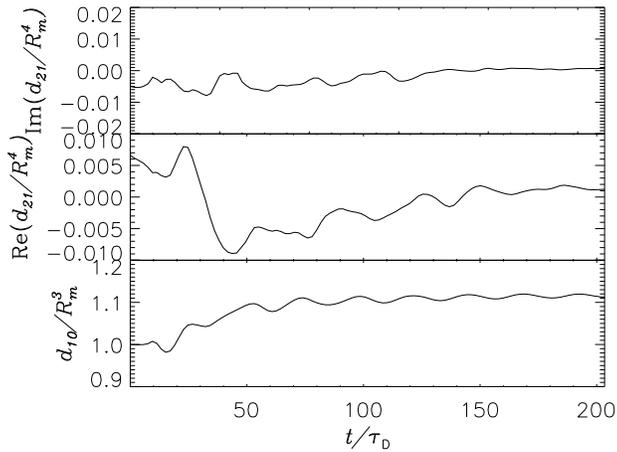}
  \caption{Magnetic dipole moment $d_{10}/R_m^3$ and magnetic quadrupole
    moment $d_{21}/R_m^4$ for model G, normalised to the initial value of
    $d_{10}/R_m^3=7.3\times10^{11}$ G, as a function of time (in units
    of the diffusion time scale $\tau_\mathrm{D}$). All other
    components of the tensor $d_{ij}$ vanish due to symmetry.}
  \label{fig:stab:modG_magmoments}
\end{figure}

The evolution of $d_{10}$, plotted in \fref{axisym:modC_magmoments},
illustrates the effect of ohmic diffusion on the magnetic
structure. Initially, $d_{10}$ is buried by the distorted
magnetic field. The resistive instability then allows $\bmath{B}$ to
straighten radially and reduce the field line curvature (cf. bottom-middle panel in
\fref{axisym:modC_2d}), as described in
\sref{axisym:general}. Ultimately, the line-tying condition of the inner boundary forces
$d_{10}$ to approach the underlying dipole moment of the star before
accretion, as the screening currents in the mountain dissipate. In
this sense, one can say that the buried magnetic field reemerges.

The physical mechanism behind reemergence is illuminated by
examining the radial dependence of the dipole moment, snapshots of
which are plotted in \fref{axisym:modC_rad_dipole}, plotted at $t/\tau_\mathrm{D}=0, 333,
665$ (solid, dotted, and dashed curves, respectively). Initially,
$d_{10}$ is screened within a thin layer near the
surface. As the screening currents dissipate resistively,
magnetic flux is transported radially outward, thereby increasing the
dipole moment measured by an outside observer.

The nonvanishing components of the magnetic dipole and 
quadrupole tensors are displayed in
\fref{stab:modG_magmoments} for an nonaxisymmetric mountain (model G).
As for the axisymmetric case (\fref{axisym:modC_magmoments}), $d_{10}$
(bottom panel) increases over the diffusion time-scale as the magnetic
field relaxes, tending to the underlying value at $r=R_\ast$. Overall,
$d_{10}$ varies by less than 10 per cent over the simulation. The
magnetic quadrupole moment $d_{21}$ approaches zero as diffusion
restores the dipolar field.

\subsection{Triaxiality}
\label{sec:nonaxisym:moments}
\begin{figure}
  \includegraphics[width=84mm, keepaspectratio]{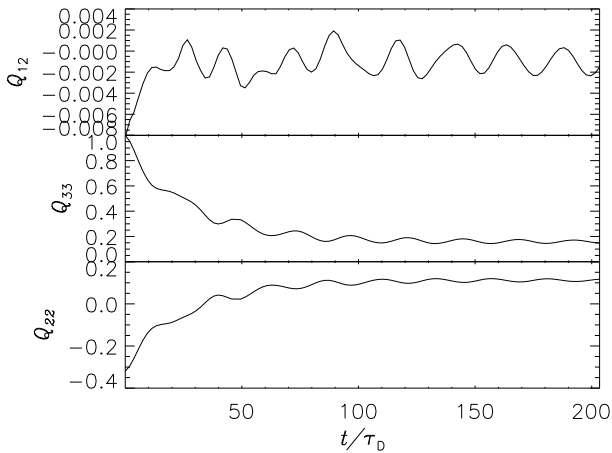}
  \caption{Components of the mass quadrupole moment tensor normalised
    to the maximum of $Q_{33}$, $7.2\times10^{24}$ g cm$^2$, as a
    function of time, in units of the diffusion time-scale, for model G.}
  \label{fig:stab:modj_quadru}
\end{figure}

The distorted magnetic field structure in \fref{axisym:modC_2d} is
accompanied by deformation of the mass distribution. This matters when
considering accreting neutron stars as gravitational wave
sources. \fref{stab:modj_quadru} plots the components of the
Cartesian mass quadrupole moment, defined as
\begin{equation}
  Q_{ij}=\int d^3 x' \, (3 x_i' x_j'-r'^2 \delta_{ij}) \rho(\bmath{x'}),
\end{equation}
versus time for the nonaxisymmetric model G. The diagonal
elements of $Q_{ij}$  measure the axisymmetric distortion and are directly
related to the ellipticity by $\epsilon \propto Q_{22} \propto
Q_{33}$. As the mountain relaxes resistively, $Q_{22}$ and $Q_{33}$
decrease, asymptoting at $\sim 20$
per cent of the initial value at $t\approx 150 \tau_\mathrm{D}$. This
is normal: plasma diffuses across the flux surfaces and spreads evenly
over the stellar surface. However, since magnetic diffusion tends to smooth out
gradients in $\bmath{B}$, thereby reducing the actual diffusion
time-scale, we reach an intermediate, metastable state. In this state, the
field lines are almost radial but the mountain has not yet
diffused to cover the surface evenly. We compute the diffusion
time-scale of the metastable state to be $\tau'_\mathrm{D}=27.6 \tau_0$,
five times higher than $\tau_\mathrm{D}=5.26 \tau_0$ of the initial
state. Eventually, the remaining plasma diffuses over the
time-scale $\tau'_\mathrm{D}$ and $Q_{ij}$ tends to zero.

The offdiagonal elements of $Q_{ij}$ (top panels of
\fref{stab:modj_quadru}) measure the deviation from
axisymmetry. They decrease on the time-scale
$\tau_\mathrm{I}$ and then oscillate 
around the abscissa. Nonaxisymmetric oscillations,
observed previously in ideal-MHD calculations (VM08), are excited
here when the mountain reconfigures: small numerical inaccuracies
perturb the steady-state equilibrium and the mountain readjusts on the
Alfv\'{e}n time-scale. The period is $\sim 10 \tau_\mathrm{D}$ for model G. The amplitude
initially grows then decays. The existence of such overstable modes is
peculiar to a dissipative MHD system. The linear force operator is no
longer self-adjoint and its eigenvalues generally have both a real and
an imaginary part. The tendency of $Q_{12}$ to decrease tallies
with the observation that resistivity restores axisymmetry by
smoothing toroidal gradients, as postulated in
\sref{nonaxisym:general}.  \change{It is important to note
  that the observed oscillations cannot arise if there is a realistic
  separation between the Alfv\'{e}n and diffusion timescale. For
  completeness, we note that the number $|Q_{11}-Q_{22}|=|Q_{33}+2
  Q_{22}|$ is another measure for the departure from
  axisymmetry. However, it is obvious from
  \fref{stab:modj_quadru} that the magnitude of this 
  number is small compared to the magnitude of the diagonal elements.}

\section{Discussion}
\label{sec:discussion}
The formation of magnetically confined mountains at the poles of an
accreting neutron star is one
explanation of the observed reduction of the magnetic dipole moment
with $M_a$ in neutron star binaries. Although a magnetic mountain is
susceptible to transient, toroidal, ideal-MHD instabilities, these are
not disruptive. The saturation state still confines the accreted
matter to the magnetic pole, efficiently screening the dipole moment
in the long term.

This article is concerned with the fate of a
magnetic mountain when a nonzero electrical resistivity switches on.
We extend the ideal-MHD code \textsc{zeus-mp} to add a resistive term
to Ohm's law and perform three-dimensional
simulations for different values of the resistivity. In the
axisymmetric case, we find that global MHD oscillations
compress the magnetic field, accelerating
plasma slippage across flux surfaces. As a consequence, the
mountain relaxes on a time-scale $\tau_\mathrm{I}$ which is shorter
than the diffusion time-scale $\tau_\mathrm{D}$ but
comparable to the accretion time-scale. In
the nonaxisymmetric case, Ohmic diffusion additionally tends to
restore axisymmetry. We do not find any evidence of transient
resistive instabilities, like the resistive ballooning mode, on the
intermediate tearing mode time-scale $(\tau_\mathrm{A}
\tau_\mathrm{D})^{1/2}$. The mountain persists over \cchange{$\sim 10^5-10^8$}
years, comparable to the duration of the accretion phase in a low-mass
X-ray binary (LMXB).

Astrophysically, the key result of the paper can be stated as follows:
\emph{magnetically confined mountains in LMXBs are stable (in ideal
  and nonideal MHD) over the accretion time-scale and relax
  over the typical life-time of radio millisecond pulsars ($\sim
  10^9$ yr) after accretion stops.} \citet{Jones04}
argued that the electrical conductivity in the solid crust is
significantly lower than that for a homogenous bcc lattice and
temperature-independent, with $\eta=10^{-24}$ s. For this value, we
expect a stationary state at $M_a \sim 10^{-5} M_\odot$ where the diffusive
mass flux escaping the polar cap is exactly replenished by
accretion. To study the structure of such a state, and confirm its
existence, we must extend the growing simulations in
\fref{grow:epsmadot}, a key topic for future work. 

One shortcoming of the calculations is the neglect of
rotation. Accreting millisecond pulsars spin up as fast as $\Omega
\sim 620$ Hz \citep{Galloway08}. \citet{Spitkovsky02} found that
surface thermonuclear burning is unaffected by rotation in its early
stages, but the thermonuclear flame spreads more slowly as time passes.
\citet{Bhattacharyya07} applied this idea to
qualitatively reproduce the light-curves from 4U 1636$-$536 and SAX
J1808.8$-$3658. The Coriolis force also modifies the continuous part
of the ideal-MHD spectrum for axisymmetric configurations with a
uniform angular velocity \citep{Hellsten79,Vigelius08d}, especially for
short-wavelength modes. \change{However, it does not affect the
  equilibrium configuration. A simple estimate shows that one requires
  a transversal speed of $v \sim 8 \times 10^8$ cm s$^{-1}$ to attain
  a Coriolis force which is comparable to $c^2 \nabla P$.} The star may
  also precess \citep{chung08}, complicating the treatment of rotational effects.

Magnetic mountains in LMXBs are promising sources of gravitational waves
\citep{Melatos05, Payne06a}. For $M_a \ga 10^{-5} M_\odot$, there is a
fair prospect of detection with next generation interferometric
detectors like the Laser Interferometric Gravitational Wave
Observatory (LIGO) (VM08). Clearly, the strength of the signal depends
critically on the long-term stability of the mountain and the rate at
which it relaxes resistively. In a companion paper
\citep{Vigelius08b}, we predict the signal-to-noise ratio attainable
by LIGO when the resistive results of this paper are included. We also
show that the electrical resistivity can be constrained by existing
LIGO data, by invoking the \citet{Bildsten98} torque-balance limit for
LMXBs and the Blandford spin-down limit for radio millisecond pulsars
\citep{Abbott07b}. 

\bibliography{resistivity.bib}

\appendix

\section{Implementing resistivity in \textsc{zeus-mp}}
\label{sec:app:resistivity}
\begin{figure*}
  \label{fig:app:resistivity:diffusion_spherical}
  \includegraphics[width=140mm,
  keepaspectratio]{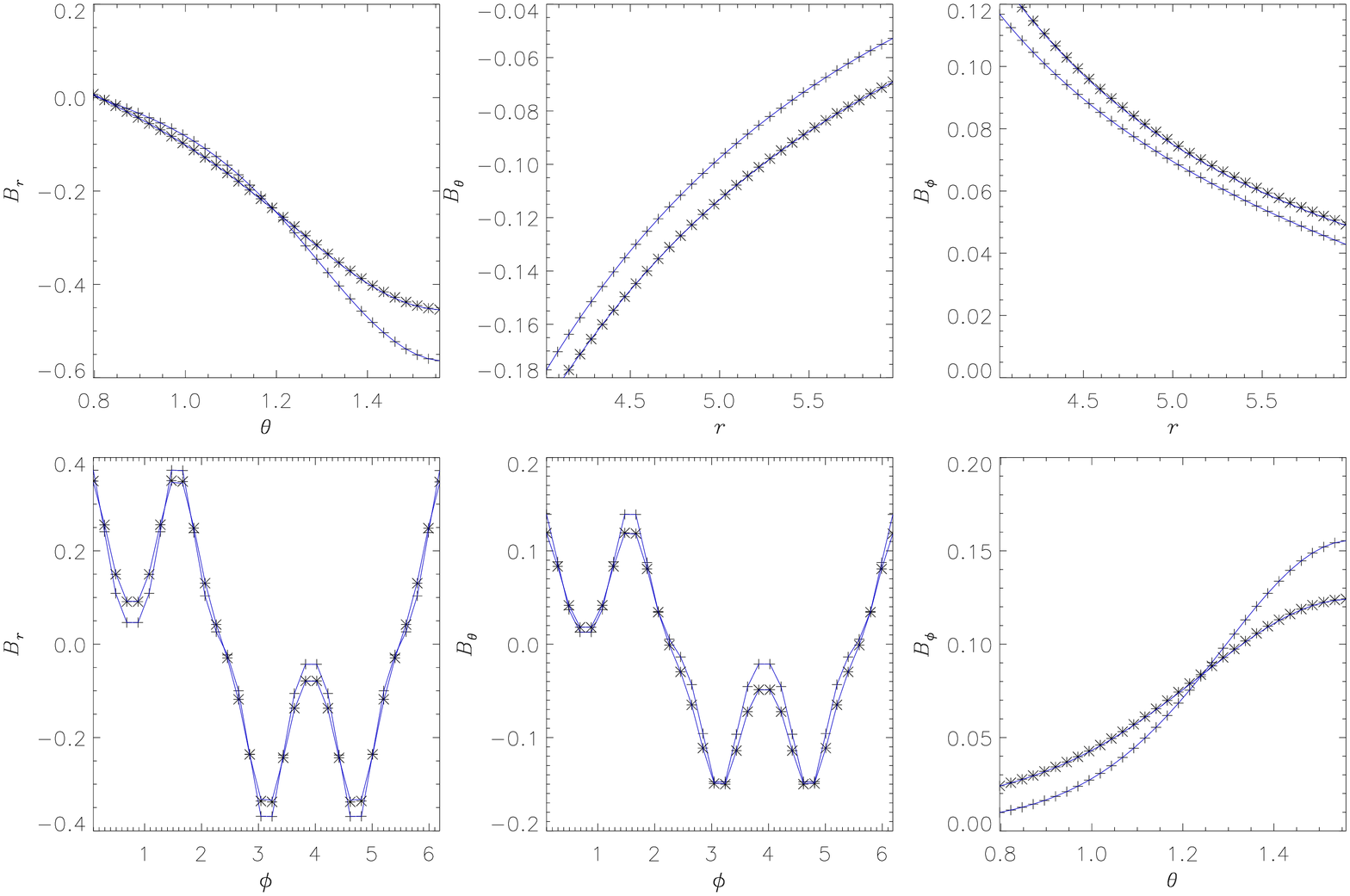}
  \caption{Snapshots of the three-dimensional diffusion problem in
    spherical coordinates at $t=\tau_0$
    (stars) and $t=\tau_0+2 \tau_d$ (crosses), where $\tau_d = 0.22 \tau_0$,
  along with the analytic solution
  \eeref{app:resistivity:analytic_1}--\eeref{app:resistivity:analytic_3}
  (solid curves).
  Shown are $B_r$, $B_\theta$, and $B_\phi$ (left, middle, and right
  panels respectively) for three cross-sections: $r=5 h_0$, $\theta=1.18$ rad,
  and $\phi=3.14$ rad. The agreement is excellent (2.4 per cent).} 
\end{figure*}

\subsection{Advection step}
\label{sec:app:resistivity:mhd}
Resistive MHD comprises a set of seven coupled, nonlinear partial
differential equations for the magnetic field $\bmath{B}$, the bulk
velocity $\bmath{v}$, the plasma density $\rho$, and the pressure $p$
\citep{Goedbloed04}: the equation of mass conservation,
\begin{equation}
  \label{eq:app:resistivity:mhd_mass_conservation}
  \frac{\partial \rho}{\partial t}+\nabla \cdot (\rho \bmath{v}) = 0,
\end{equation}
the momentum equation,
\begin{equation}
  \label{eq:app:resistivity:mhd_momentum}
  \rho \left(\frac{\partial \bmath{v}}{\partial t} + \bmath{v}\cdot
    \nabla \bmath{v} \right) + \nabla p -  (\nabla \times
  \bmath{B}) \times \bmath{B}+\rho \nabla \varphi=0,
\end{equation}
and the induction equation,
\begin{equation}
  \label{eq:app:resistivity:mhd_induction}
  \frac{\partial \bmath{B}}{\partial t} - \nabla \times
  (\bmath{v}\times \bmath{B} - \eta \nabla \times \bmath{B}) = 0,
\end{equation}
where $\eta$ denotes the resistivity and $\varphi$ is the
gravitational field.
The system is closed by the supplementary condition $\nabla \cdot
\bmath{B}=0$ and an isothermal equation of state $p=c_s^2 \rho$,
where $c_s$ represents the isothermal sound speed. 

As explained by \citet{Hayes06}, \textsc{zeus-mp} employs an
operator split algorithm based on the method of finite differences on
a staggered grid. The advection step is done in two stages. Firstly, a
source step solves 
\begin{equation}
  \rho \frac{\partial \bmath{v}}{\partial t}=-\nabla p - \nabla \cdot
  \bmath{Q} - \rho \nabla \varphi - \nabla (B^2/2 \mu_0),
\end{equation}
where an artificial viscous pressure tensor $\bmath{Q}$ is included
\citep{vonNeumann50}. Secondly,
to treat transversal MHD waves properly, one must solve the magnetic
tension force along with the induction equation in a single step using
the method of characteristics and constrained transport (MOCCT)
\citep{Hawley95}.
The MOCCT step advances $\bmath{B}$ by computing
the line integral of the electromotive force (EMF)
$\bepsilon=\bmath{v}\times\bmath{B}$
around a cell boundary $S$:
\begin{equation}
  \label{eq:app:resistivity:mocct}
  \frac{\mathrm{d}}{\mathrm{d} t} \int_S \bmath{B}\cdot \mathrm{d}\bmath{S}=\oint_{\partial S}
  \bepsilon\cdot \mathrm{d}\bmath{l}.
\end{equation}
Second-order accuracy in time is achieved by employing time-centered values for
$\epsilon$. The extrapolation in time is done using the characteristic
equation for transverse Alfv\'{e}n waves. It can be shown that MOCCT ensures $\nabla \cdot \bmath{B}=0$
to machine accuracy at all times provided the initial field is
solenoidal. The extrapolated $\bmath{B}$ is then used to work out the transverse
magnetic forces and accelerate the fluid accordingly:
\begin{equation}
  \rho \left.\frac{\partial \bmath{v}}{\partial
      t}\right|_\mathrm{final} = \left.\frac{\partial \bmath{v}}{\partial
      t}\right|_\mathrm{source step}+\mu_0^{-1} (\bmath{B}\cdot\nabla)\bmath{B}. 
\end{equation}
Finally, the fluid density and momentum are advected via
\begin{equation}
  \frac{\mathrm{d}}{\mathrm{d} t} \int_V \rho \; \mathrm{d}V = - \oint_{\partial V} \rho
  \bmath{v} \cdot \mathrm{d} \bmath{S}
\end{equation}
and
\begin{equation}
  \frac{d}{\mathrm{d} t} \int_V \rho \bmath{v} \; \mathrm{d}V = - \oint_{\partial V} \rho
  \bmath{v} \bmath{v} \cdot \mathrm{d} \bmath{S}.
\end{equation}

A visual comparison of \eeref{app:resistivity:mhd_induction} and
\eeref{app:resistivity:mocct} suggests a natural way to incorporate the
resistive term \citep{Stone99}: we use the updated $\bmath{B}$
to work out the current density $\bmath{j}= \nabla \times \bmath{B}$
and apply \eref{app:resistivity:mocct} again, replacing
$\bepsilon$ by $-\eta \bmath{j}$ this time. The staggered grid allows
for central differencing and thereby guarantees spatial second-order
accuracy. This resistive algorithm has been
used in conjunction with \textsc{zeus-3d} to study
protostellar jet formation \citep{Fendt02}.

A von Neumann analysis in terms of eigenmodes yields a stability
criterion for parabolic PDEs \citep{Press86}, $\Delta t \le 2 \Delta^2
\eta^{-1}$, which depends quadratically on the minimal grid cell size
$\Delta$. We find empirically that our implementation requires
$\Delta t \le 10^{-2} \Delta^2 \eta^{-1}$. If $\eta$ is high, this
constraint dominates the ideal-MHD timestep and drastically increases the run
time. We therefore make use of a superstep algorithm, similar to the
one described by \citet{Alexiades96}. We compute the ideal-MHD
timestep $\Delta t_\mathrm{CFL}$ according to the usual
Courant-Friedrichs-Levy (CFL) condition, as well as the resistive 
timestep $\Delta t_\mathrm{resistive}$. After updating $\bmath{B}$ by the MOCCT procedure, we apply
the resistive algorithm in a cycle of $N$ steps, such that $\Delta
t_\mathrm{CFL} = N \Delta T$, with $\Delta T \le \Delta
t_\mathrm{resistive}$. This approach was implemented successfully in a
resistive module for the \textsc{pluto} code \citep{Mignone07}.

Special care must be taken when incorporating the boundary conditions.
 \textsc{zeus-mp} adds two and three ghost cells at
the inner and outer boundaries, respectively, where it either sets
$\bepsilon$ according to the boundary conditions for $\bmath{B}$ and
$\bmath{v}$ or communicates it at a processor boundary (inside the integration volume) via the message
passing interface (MPI). The processor boundaries are set by the MPI
topology, i.e. the division of the computation grid among the different
processors. The staggered grid requires $\bepsilon$ at the
inner boundary, so we need to add another layer of
ghost cells for $\bmath{B}$ and $\bmath{v}$ there to provide
$\bepsilon$ in all ghost cells. In order to minimize the alterations to the
code, we prefer to compute (or communicate to the neighbouring processor) the
whole layer of ghost cells for $\bmath{B}$ at the beginning of every
super-step. We employ the MPI communication flow described in
\citet{Hayes06} to minimize inter-processor traffic.

\subsection{Test case}
\label{sec:app:resistivity:test}

We test our code extensions by simulating a purely diffusive
problem. We set $\rho=10^9 \rho_0$ (see \sref{model:units}) and $\bmath{v}=0$ to suppress any
fluid motions. Equations \eeref{app:resistivity:mhd_mass_conservation}--\eeref{app:resistivity:mhd_induction}
then reduce to a single diffusion equation
\begin{equation}
  \label{eq:app:resistivity:diffusion}
  \frac{\partial \bmath{B}}{\partial t}=-\nabla\times(\eta \nabla
  \times \bmath{B}),
\end{equation}
which can be solved easily in Cartesian coordinates $(x,y,z)$:
\begin{eqnarray}
  \bmath{B}(\bmath{r},t)&=&\frac{\mathrm{e}^{-4 \eta t}}{t}
  \\
  & & \times [\mathrm{e}^{-(y^2+z^2)} \unitx + \mathrm{e}^{-(x^2+z^2)} \unity +
    \mathrm{e}^{-(x^2+y^2)} \unitz]. \nonumber
\end{eqnarray}
A coordinate transformation yields the result in spherical coordinates:
\begin{eqnarray}
  \label{eq:app:resistivity:analytic_1}
  B_r (\bmath{r},t) & = & \frac{\mathrm{e}^\frac{-r^2}{4 \eta t}}{t}
  \left[e^{\frac{r^2 \cos^2 \theta}{4 \eta t}} \cos \theta +\sin \theta \right. \\
 & & \times \left. 
  \left(e^{\frac{r^2 \cos^2 \phi \sin^2 \theta}{4 \eta t}} \cos \phi+
    e^{\frac{r^2 \sin^2\theta \sin^2\phi}{4\eta t }} \sin \phi
  \right)\right],  \nonumber
\end{eqnarray}
\begin{eqnarray}
  B_\theta (\bmath{r},t) & = & \frac{e^{-\frac{r^2}{4 \eta t}}}{t}
  \left[\cos \theta \right. \\
    & & \times \left(e^{\frac{r^2 \cos^2\phi \sin^2 \theta}{4 \eta t}} \cos \phi
    +e^{\frac{r^2 \sin^2 \theta \sin^2 \phi}{4 \eta t }} \sin \phi
\right) 
      \nonumber \\
  & & \left.  -e^{\frac{r^2 \cos^2 \theta}{4 \eta t }} \sin \theta \right],
\end{eqnarray}
and
\begin{eqnarray}
  \label{eq:app:resistivity:analytic_3}
  B_\phi (\bmath{r},t) & = & \frac{e^{-\frac{r^2}{4 \eta t }}}{t}
    \\ & & \times \left(e^{\frac{r^2 \sin^2 \theta \sin^2 \phi}{4 \eta t }}
      \cos \phi -e^{\frac{r^2 \cos^2 \phi \sin^2 \theta}{4 \eta t }}
      \sin \phi \right). \nonumber
\end{eqnarray}
Test runs were performed in Cartesian (three dimensions)
and spherical polar coordinates (two and
three dimensions) employing time-dependent boundary
conditions with Eqs. \eeref{app:resistivity:analytic_1} and
\eeref{app:resistivity:analytic_3}. \fref{app:resistivity:diffusion_spherical} shows
two snaphots of 
the three-dimensional run in spherical polar coordinates at $t=\tau_0$
(stars) and $t=\tau_0+2 \tau_d$ (crosses), with $t_d = 0.22 \tau_0$,
along with the analytic solution
\eeref{app:resistivity:analytic_1}--\eeref{app:resistivity:analytic_3}.
They are in excellent agreement, with  relative error $< 2.4$
per cent at $t=\tau_0+2 \tau_\mathrm{d}$.

\section{Resistive Parker instability}
\label{sec:app:parker}
In this section, we derive an analytic dispersion relation for the linear,
resistive, MHD modes of a plane-parallel, gravitating plasma
slab \citep{Singh69}. The ideal-MHD counterpart of this problem is known as the Parker
instability \citep{Parker1967, Mouschovias74}.

Let us assume a uniform gravitational acceleration $g$ directed parallel to the
$z$-axis, and a unidirectional magnetic field parallel to the
$y$-axis. We can then write down the magnetostatic equilibrium.
The density and magnetic field are given by
\begin{equation}
  \rho(z)=\rho_0 \exp \left[\frac{-g z}{u^2 (1+\alpha)}\right],
\end{equation}
\begin{equation}
  \bmath{B}(z) = B_0 \exp \left[\frac{-g z}{2 u^2 (1+\alpha)}\right] \bmath{e}_y,
\end{equation}
under the additional assumption that the magnetic pressure is
proportional to the gas pressure everywhere, viz.
\begin{equation}
  \frac{B^2}{2 \mu_0} = \alpha p.
\end{equation}
The equation of state is $p=u^2 \rho$.

Next, we write down the linearized equations of mass conservation,
\begin{equation}
  \label{eq:parker:mass}
  \frac{\partial \drho1}{\partial t}+\rho \nabla \bcdot \V1 + \V1 \bcdot
  \nabla \rho =0,
\end{equation}
force balance,
\begin{eqnarray}
  \label{eq:parker:force}
  \rho \frac{\partial \V1}{\partial t}&=&-\nabla \p1 - \frac{1}{2 \mu_0}
  \nabla \left[2 \bmath{B} \bcdot \B1 \right] \\
  & & + \frac{1}{\mu_0} \bmath{B}
  \bcdot \nabla \B1 + \frac{1}{\mu_0} \B1
  \bcdot \nabla \bmath{B} - \drho1 g \bmath{e}_z,
\end{eqnarray}
and induction,
\begin{equation}
  \label{eq:parker:induction}
  \frac{\partial \B1}{\partial t}=\frac{c^2}{4 \pi \sigma}
  \nabla^2 \B1 + \nabla \times \left[ \V1 \times \bmath{B} \right].
\end{equation}
In \eeref{parker:mass}--\eeref{parker:induction} and below, $\p1$,
$\drho1$, $\B1$, and $\V1$ denote the 
perturbations of the pressure, density,  magnetic field, and 
velocity respectively. $\sigma$ denotes the conductivity.

Ignoring interchange modes ($k_x=0$), we assume the perturbed
quantities have the form $\propto \mathrm{e}^{\i (k_y y-\omega
  t)}$. Furthermore, we only consider perturbations in the $y$-$z$
  plane, i.e. $\dv1_x=0$ and $\B1=[0, \partial \a1_x/\partial z, -\partial \a1_x/\partial
    y]$, where $\bmath{A}=A_x \bmath{\hat{x}}$ is the vector
    potential. Eq. \eeref{parker:mass} yields 
  \begin{equation}
    \label{eq:parker:perturb1}
    -\i \omega \drho1 + \i k_y \rho \dv1_y+\rho \frac{\partial
      \dv1_z}{\partial z}-\frac{1}{L} \dv1_z \rho =0,
  \end{equation}
with $L=u^2(1+\alpha)/g$. Similarly, the components of
\eeref{parker:force} reduce to
\begin{equation}
    \label{eq:parker:perturb2}
  -\i \omega p \dv1_y = -\i u^2 k_y \p1 + \b1_z \frac{\alpha
    u^2\rho}{L B_y},
\end{equation}
and
\begin{eqnarray}
    \label{eq:parker:perturb3}
  -\i \omega \rho \dv1_z & = & -u^2 \frac{\partial \p1}{\partial z}
  - \frac{\partial \a1_x}{\partial z} \frac{\alpha u^2 \rho}{L B_y} \\
 & & -\frac{1}{\mu_0} B_y \left[ \frac{\partial^2 \a1_x}{\partial
     z^2}+ \frac{\partial^2 \a1_x}{\partial y^2}\right] - \drho1 g.
\end{eqnarray}
Finally, the induction equation \eeref{parker:induction} yields
\begin{equation}
   \label{eq:parker:perturb4}
  \frac{\partial \a1_x}{\partial t} = \frac{c^2}{4 \pi \sigma}
  \left(-k_y^2+\frac{\partial^2}{\partial z^2} \right) \a1_x - B_y \dv1_z.
\end{equation}

In order to solve \eeref{parker:perturb1}--\eeref{parker:perturb4}
analytically, we make the short-wavelength approximation  $\partial/\partial z \ll
k_y$. Eliminating $\dv1_z$ and $\drho1$, we find the dispersion relation
\begin{equation}
  \label{eq:parker:dispersion}
  \left(\omega^2+\frac{\i \omega}{\mu_0 \sigma} k_y^2-2 \alpha u^2
    k_y^2 \right)\left(\omega^2-u^2 k_y^2\right) = \frac{g^2 \alpha}{1+\alpha}.
\end{equation}
For $\sigma \rightarrow \infty$, \eeref{parker:dispersion} can be
solved to obtain
\begin{equation}
  2 \omega_\infty^2 = (1-2 \alpha)u^2 k_y^2 \pm \left[ (1-2 \alpha)^2
    u^4 k_y^4 - \frac{4 k_y^2 g^2 \alpha}{1+\alpha}\right]^{1/2}.
\end{equation}
The modes are stable when the discriminant is positive:
\begin{equation}
  \frac{u^4 k_y^4}{g^2} > \frac{4 \alpha}{1+\alpha} \frac{1}{(1-2 \alpha)^2}.
\end{equation}
For large but finite $\sigma$, $\omega_\infty$ is perturbed slightly,
with $\omega=\omega_\infty+\omega'$ and $|\omega'|\ll
\omega_\infty$. Solving for $\omega'$, we obtain two branches, the
first damped,
\begin{equation}
  \omega'_\mathrm{damped}=-\frac{\i k_y^2 c^2}{8 \pi \sigma} = -
  \frac{\i}{2 \tau_\mathrm{D}}
\end{equation}
and the second growing,
\begin{eqnarray}
  \label{eq:parker:growing}
  \omega'_\mathrm{growing}& = & \frac{\i c^2}{16 \pi \sigma} \frac{2 \alpha}{1+\alpha}
        \frac{g^2}{(1-2 \alpha)^2 u^4}\\
        &= &\frac{1}{2 (1-2 \alpha)^2}
        \frac{\alpha}{1+\alpha} \frac{g^2}{k_y^2 u^2} \frac{\i}{\tau_\mathrm{D}}.
\end{eqnarray}
The damped mode has a \change{decay} time roughly equal to the diffusion
time $\tau_\mathrm{D}$, whereas the growing mode amplifies quickly,
over the time required to diffuse across one hydrostatic scale
height.

\end{document}